\documentclass[10pt,conference]{IEEEtran}
\IEEEoverridecommandlockouts
\usepackage{cite}
\usepackage{amsmath,amssymb,amsfonts}
\usepackage{algorithmic}
\usepackage{graphicx}
\usepackage{textcomp}
\usepackage{xcolor}

\usepackage{enumitem}
\usepackage{todonotes}
\usepackage{csquotes}
\usepackage{multicol}
\usepackage{multirow}
\usepackage{balance}
\usepackage{subfig}
\usepackage{booktabs}

\usepackage[numbers]{natbib}

\usepackage{color, colortbl}
\definecolor{bostonuniversityred}{rgb}{0.8, 0.0, 0.0}
\definecolor{medblue}{rgb}{0.0, 0.0, 0.7}
\definecolor{light-gray}{gray}{0.95}

\newcommand{\red}[1]{#1}

\definecolor{light-cyan}{rgb}{0.88,1,1}
\definecolor{table-highlight}{gray}{0.95}

\def\BibTeX{{\rm B\kern-.05em{\sc i\kern-.025em b}\kern-.08em
    T\kern-.1667em\lower.7ex\hbox{E}\kern-.125emX}}
\begin{document}

\title{Morning or Evening? An Examination of Circadian Rhythms of CS1 Students}

\makeatletter
\newcommand{\linebreakand}{%
  \end{@IEEEauthorhalign}
  \hfill\mbox{}\par
  \mbox{}\hfill\begin{@IEEEauthorhalign}
}
\makeatother

\author{
\IEEEauthorblockN{Albina Zavgorodniaia}
\IEEEauthorblockA{\textit{Computer Science Department} \\
\textit{Aalto University}\\
Espoo, Finland \\
albina.zavgorodniaia@aalto.fi}
\and
\IEEEauthorblockN{Raj Shrestha}
\IEEEauthorblockA{\textit{Computer Science Department} \\
\textit{Utah State University}\\
Logan, Utah, USA \\
raj.shrestha86@outlook.com}
\and
\IEEEauthorblockN{Juho Leinonen}
\IEEEauthorblockA{\textit{Computer Science Department} \\
\textit{University of Helsinki}\\
Helsinki, Finland \\
juho.leinonen@helsinki.fi}
\linebreakand
\IEEEauthorblockN{Arto Hellas}
\IEEEauthorblockA{\textit{Computer Science Department} \\
\textit{Aalto University}\\
Espoo, Finland \\
arto.hellas@aalto.fi}
\and
\IEEEauthorblockN{John Edwards}
\IEEEauthorblockA{\textit{Computer Science Department} \\
\textit{Utah State University}\\
Logan, Utah, USA \\
john.edwards@usu.edu}
}

\maketitle

\begin{abstract}

Circadian rhythms are the cycles of our internal clock that play a key role in governing when we sleep and when we are active. A related concept is chronotype, which is a person's natural tendency toward activity at certain times of day and typically governs when the individual is most alert and productive. In this work we investigate chronotypes in the setting of an Introductory Computer Programming (CS1) course. Using keystroke data collected from students we investigate the existence of chronotypes through unsupervised learning. The chronotypes we find align with those of typical populations reported in the literature and our results support correlations of certain chronotypes to academic achievement. We also find a lack of support for the still-popular stereotype of a computer programmer as a night owl. The analyses are conducted on data from two universities, one in the US and one in Europe, that use different teaching methods. In comparison of the two contexts, we look into programming assignment design and administration that may promote better programming practices among students in terms of procrastination and effort.

\end{abstract}





\section{Introduction}

Circadian rhythms -- the cycles of our internal clock -- influence our daily activity and productivity. Achievements in academic studies~\cite{preckel2011chronotype}, programming~\cite{Fucci2020}, and other domains are affected by circadian rhythms. Chronotype, or diurnal preference, is a person's tendency toward activity at certain times of day and is thought to be a natural characteristic of an individual~\cite{Roenneberg2003}.

Gaining an understanding of students' internal clocks and time-management behaviors has the potential to inform individualized recommendations for learning as well as targeted interventions for helping those struggling with managing their time.
With that goal, we study introductory programming students from two contexts, one in the US and one in Europe, with different geographical location leading to differences (e.g. available daylight). In addition, the contexts have different teaching approaches and use different programming languages. In both contexts, we collected timestamped keystroke data from students' programming process for the purposes of gaining a deeper understanding of students' behavior. Prior work in computing education has used keystroke data, for example, to predict students' success~\cite{leinonen2016automatic,edwards2020study,leinonen2019keystroke} and to identify students taking an exam~\cite{longi2015identification,leinonen2016typing}. Our overarching objective is to study the students' possible chronotypes and preferences, as evidenced by students' behavior inferred from the timestamped keystrokes, and look for connections of these chronotypes and preferences with assessment outcomes. 

In the context of software engineering education and computing education research, there exist streams of research on students' time management practices and self-regulation~\cite{bergin2005influence,leppanen2016pauses,ilves2018supporting}. Much of this has had a focus on behavior related to deadlines, including procrastination, and ways to increase the earliness of students' work~\cite{edwards2009comparing,martin2015effects,ilves2018supporting}. In general, these studies agree that starting early leads to better learning outcomes, but little focus has been put on at what times students typically work and how these times contribute to students' learning outcomes, despite some existing research on software developers' working hours and bugs introduced in code commits made at different times during the day~\cite{Claes2018,Fucci2020} . 

Our work offers three contributions. 1) We use unsupervised machine learning to identify chronotypes among introductory computer programming (CS1) students and find that these chronotypes match up remarkably well with chronotypes of general populations reported in the literature. 2) We find strong linkages between chronotype and academic outcomes, again, consistent with the literature. We also highlight circadian rhythms as a viable, important, and relevant topic for both the computing education research community as well as for computing educators.

This article is organized as follows. Next, in Section~\ref{sec:background}, we outline the theory and related work upon which our article builds. Section~\ref{sec:methodology} describes the contexts in which the study took place and outlines the research questions and methodology. Results are presented in Section~\ref{sec:results}, and discussed in Section~\ref{sec:discussion}. Finally, conclusion and potential streams for future work are outlined in Section~\ref{sec:conclusions}.

\section{Background and Related Work \label{sec:background}}


\subsection{Circadian Rhythms and Chronotypes}

A \emph{circadian rhythm} is a cycle of internal oscillations in nearly all physiological activities generated by the molecular circadian clock and has a period of approximately 24 hours \cite{Reppert2002,Dibner2010}. \emph{Chronotype} is a person's natural inclination toward activity at certain periods of the day and depends on a circadian rhythm for synchronization~\cite{Roenneberg2003}.
Research suggests that the circadian cycle is conditioned by a group of clock genes \cite{BellPedersen2005}, which explains individual differences. Nevertheless, it is not fixed and it does shift during an individual's lifetime. 

People have to fit in daily activity according to their schedules taking into consideration social constraints (e.g., typical work times, services work hours, etc.). That is, clock time preferences are more likely to match a chronotype when an individual has more flexibility in activity organisation. Discrepancy between an individual's chronotype and schedules determined by social constraints causes a phenomenon called \emph{social jetlag}~\cite{Roenneberg2003,Wittmann2006}. It leads to accumulation of a ''sleep debt'' with subsequent feelings of tiredness and drop in cognitive abilities throughout a working week. Research \cite{Korczak2008,Vitale2014} shows that sleep-wake times of people whose weekday routine matches their chronotype do not change during weekends contrary to the case of people whose timing inclination of natural activity contrast with their weekday routine. Since both activity and sleep comply with chronotype in natural conditions, there is a premise to consider weekends as a reflection of true chronotypes.  

Most studies distinguish between two chronotypes: \textit{morning} and \textit{evening}. In recent years a third type -- \textit{intermediate} has emerged (for example, in \cite{Valladares2017,Porcheret2018}).  In 2019, Putilov et al. \cite{Putilov2019} formed an argument that there are four chronotypes which can be distinguished by varying times of sleepiness and alertness: \textit{morning}, \textit{afternoon}, \textit{napper} and \textit{evening}. The \emph{\textbf{morning}} type prefers to wake up early in the morning and is most alert from 9 a.m. to 11 a.m., after which the alertness curve gradually goes down. The \emph{\textbf{napper}} type has two peaks of activity -- the first from 9 a.m. to 11 a.m., and the second from 3 p.m. to 10 p.m. The \emph{\textbf{afternoon}} type is alert from 11 a.m. to 6 p.m., gradually decreasing until 9 p.m. 
Finally, the \emph{\textbf{evening}} type, similarly to the napper type, has two peaks, although the first one is less evident and falls into period of 11 a.m. to 2 p.m., while the second period of increased alertness is from 6 p.m. until late night. We use Putilov's et al. chronotype classification in our study.

Relationships between chronotype and academic achievement of adolescents have been discussed across countries (e.g., \cite{Kolomeichuk2016,Valladares2017,Montaruli2019,dunn1987research,dunn1988learning}). There exist consistent patterns of negative correlation between ``eveningness'' and indicators of academic performance (e.g. GPA, exam grades), whereas ``morningness'' is positively correlated with academic achievement. It is possible that these correlations are at least partly due to the \textit{synchrony effect}~\cite{nowack2018synchrony}, where an individual's cognitive performance is optimal during certain times of day and suboptimal at other times. That is, a student working during hours that are suboptimal for them because of socially driven schooltime hours~\cite{Roenneberg2003,Wittmann2006} can lead to sub-par academic achievement.

Attention, working memory, and executive functions are affected by the synchrony effect~\cite{Cajochen1999,Schmidt2007,Valdez2008}, as well as the ability to inhibit distractors and focus on a specific task~\cite{Hasher2008}. Furthermore, the synchrony effect is more pronounced when tasks are more difficult~\cite{Natale1997, Natale2003}.

\subsection{Chronotypes in Software Engineering}
Recent research by~\citet{Claes2018} examined software developers' code contributions to a large number of open source projects. The study showed that most (2/3) of the developers followed typical working hours. A small cluster of developers tended to work outside the regular daily schedule and also contributed on weekends. Although those who worked late hours and weekends could be explained by the ``evening'' chronotype of the developers, an alternative explanation is that, since the study used open source projects many of which pay little or no salary to contributors, developers were working another job during normal working hours. The times of code contributions have also been linked with occurrence of bugs~\cite{Fucci2020}, providing potential evidence of sleep deprivation and inconsistency with individual circadian rhythms. 

Time management among students has received attention in computing education research. Measuring time usage or time management, by itself, can be challenging as different approaches for measuring time use (e.g. self-reported time, time logged from a learning environment) may not correlate strongly~\cite{leinonen2017comparison}. It has been suggested that surveys used for studying students' learning behaviors such as the MSLQ~\cite{pintrich1991manual} may measure what students think they do instead of what they actually do~\cite{leppanen2016pauses}, even though some of the metrics correlate with course success~\cite{bergin2005influence}. Despite the difficulty in measuring time usage, it has been shown that the amount of time that students spend on exercises tends to have an effect on exercise scores, in general~\cite{spacco2015analyzing,leinonen2017comparison}, and the way how students space their time when working on assignments can contribute to course outcomes~\cite{leppanen2016pauses}. One of the concerns related to students' time management behavior is procrastination, i.e., students delaying the start of working on course projects despite knowing that it might lead to worse outcomes. When studying student data over multiple years,~\citet{edwards2009comparing} observed that students starting their project work late in general tend to earn poorer grades than those starting early.

Researchers have shown various ways that can influence when students start their work. These include procrastination interventions such as e-mail alerts~\cite{martin2015effects}, dashboards or visualizations that illustrate how students are managing their time and may implicitly set objectives on time management to students~\cite{auvinen2015increasing,ilves2018supporting}, and more generic course design principles, such as offering practice tasks before larger projects, which are easier to start and consequently led to course work being started earlier~\cite{denny2018improving}.

As shown, students' use of time has received some attention from the computing education research domain, but time management has not been studied extensively from the theory perspective~\cite{prather2020we}, and, to our knowledge, there has been no work on student chronotypes in the computing education research context.

\section{Methodology \label{sec:methodology}}

\subsection{Context and Data}
We collected keystroke data in two contexts. For simplicity we refer to the contexts as \textit{US} and \textit{European}. However, as we describe below, differences between the contexts are more far-reaching than just the geographic location. Differences in the contexts are summarized in Table~\ref{tab:ctx-summary}.

\subsubsection{US}
Keystroke data was collected during the first five weeks of a Python-based CS1 course at a mid-sized public university in the US over the course of two semesters. The custom IDE logged timestamped keystrokes from weekly programming projects. After the five weeks students transitioned to a mainstream Python IDE that did not support keystroke logging capability. 525 students participated in the study. There were three sections of the course each semester, all taught by the same instructor except for one section in the first semester. All classes were held on Mondays, Wednesdays, and Fridays. Students were required to attend a weekly recitation where students met in smaller groups with a Teaching Assistant. During the five weeks there were five programming assignments (A1,A2,...,A5), one due each Friday at midnight. Assignments could be turned in up to one day late for a 25\% penalty and each student could turn in one assignment up to two days late without penalty. The assignments were a mix of standard text-based programming projects (e.g. mortgage calculator) and turtle-based graphics projects (e.g. animate a dartboard game). The projects were manually assessed and were each worth 100 points. After three weeks a first midterm exam was administered with a second midterm exam given four weeks later. The exams were administered on a computer in a testing center and included multiple choice and true/false questions. Some simple fill in the blank questions were included, such as, ``what does the following code output to the screen?''

\subsubsection{European}

Keystroke data was collected for the duration of a 7-week Java-based CS1 course at a large public research-first university in Finland. The course follows a pedagogy where students work on tens of programming exercises each week, including exercises with multiple graded steps realized through intermediate goals (or ``subgoals''~\cite{margulieux2012subgoal}). Exercises and course materials were released weekly, and students had approximately one week to complete the exercises for a particular week. The exercise handouts were embedded in the course materials so that whenever a new topic was learned, students immediately saw programming exercises that they could work on to internalize the topic. 

Exercises were worked on in a custom desktop IDE that kept track of the exercise that the student was working on and provided support for testing and submitting the exercises. In addition, the IDE collected timestamped keystrokes, used for both detecting plagiarism~\cite{leinonen2016typing,hellas2017plagiarism} and for course improvement and research purposes. The exercises were automatically assessed and submitting work for grading after the deadline was not possible. 

The course under study had one weekly lecture, held on Tuesdays, and walk-in labs, which were open daily. In the walk-in labs, students were guided by the course teacher and course assistants. There were two exams, one in the middle of the course and one at the end of the course. The exams were computer-based and contained programming tasks similar to the ones that students had worked on during the course. Grading-wise, 70\% of the course score comes from completed exercises, and 30\% from the exams. 

For the analysis conducted in this article, we focus on students who attended at least two weeks of the course. This is due to the university allowing sampling courses (for no fee), and withdrawing with no repercussions.

\begin{table}[t!]
  \centering
  \caption{Summary of contexts. Daylight is measured on January 1.}
  \begin{tabular}{lcc}
    \toprule
    & US & European \\
    \midrule
    Daylight (hrs) & 9 & 6 \\
    Language (prog.)  & Python       & Java          \\
    Language (inst.)  & English      & Finnish       \\
    Participants     & \red{519}          & 318       \\
    \# keystrokes     & 11,185,716   & 20,081,066 \\
    Due dates         & flexible           & fixed    \\
    Projects         & 10           & 147    \\
    Exam         & Midterm          & Midterm \& Final          \\
  \bottomrule
\end{tabular}
  \label{tab:ctx-summary}
\end{table}

\subsection{Research Questions}
Our research questions for this study are as follows.

\begin{itemize}

    \item[\bf{RQ1}] What chronotypes would be discovered from clustering keystroke data collected from two contexts?
    \item[\bf{RQ2}] How do the typical working times of students relate to academic outcomes?
    \item[\bf{RQ3}] How do contextual factors affect academic outcomes?

\end{itemize}

\subsection{Metrics}

We analyze distributions of keystroke timestamps. In order to identify student chronotypes from the data we first assign a 4-dimensional feature vector to each student, where the vector is the percentage of the student's keystrokes that occurred in different ranges of hours: [3-9, 9-15, 15-21, 21-3]. So a student with a feature vector of, say, [0.08, 0.1, 0.19, 0.63] types $63\%$ of their keystrokes between 21:00 and 3:00, so they would likely be classified as having an evening or night chronotype. After computing feature vectors for each student we used $k$-means clustering to identify student chronotypes. Each of the two contexts (US and European) were clustered separately. We used the elbow method to decide $k$.
Two reasonable candidate values of $k$ for the US context were 4 and 8, while for European, 3 and 5 were the best. We chose to use $k=4$ in order to have a consistent value for both the US and European contexts and because it allowed us to compare our results directly with those of~\citet{Putilov2019}, who proposed four chronotypes. After identifying clusters of students, we study distributions of course outcomes per identified chronotype, including assignment scores, exam scores, hours before deadline, and number of keystrokes.

The hour ranges for the feature vector are chosen arbitrarily. As such, we must take care to ensure that this particular selection doesn't introduce bias into the analysis. Ideally we would bin keystrokes into 24 one-hour bins, but clustering on 24-dimensional vectors suffers from the curse of dimensionality. Instead, we bin student keystrokes twice more using shifted hour ranges of [5-11, 11-17, 17-23, 23-5] and [7-13, 13-19, 19-1, 1-7]. We then cluster students based on these feature vectors and compare the resulting clusters across the three binning strategies. Similarity of distributions indicates robustness to binning strategy whereas distributions that are not similar indicate a bias caused by the selected bins.

All programming project keystrokes during the study period in the US context were used in our analyses. In the European context, when analyzing the hours before deadline, we excluded keystrokes from weeks 4 and 7. During week 4 there were technical issues with the assessment system and the weekly deadlines were prolonged, while in week 7 students had two weeks for completing the assignments instead of the normal one week.

In our analysis we performed 16 statistical significance tests, 8 of them Kruskal-Wallis H, 4 Mann-Whitney U, and 4 chi-squared. We report $p$-values of our statistical tests as one component among others that together contribute to understanding of our results~\cite{wasserstein2016asa}. We do not make threshold-based claims of statistical significance~\cite{carver1993case}.

\subsection{Ethics}
\red{Data from the United States was collected and analyzed under Utah State University IRB 9580. Data from the European context was collected and analyzed with student consent according to the ethical protocols outlined by The Finnish National Board on Research Integrity TENK\footnote{https://tenk.fi/en/ethical-review/ethical-review-human-sciences}.}

\section{Results \label{sec:results}}

\begin{figure*}
\centering
    \subfloat[][Morning]{
    \includegraphics[width=0.83\columnwidth]{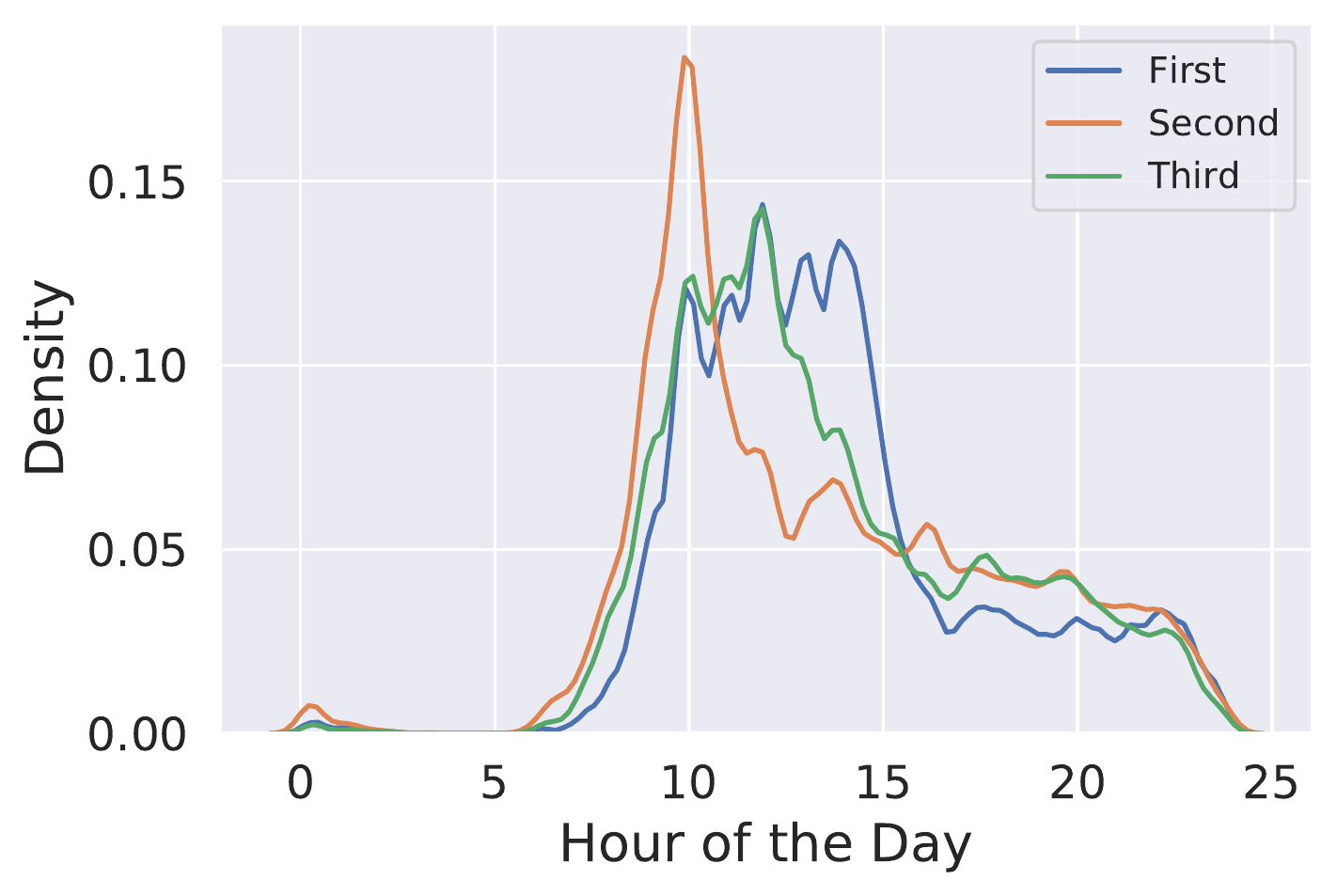}
    }
    \subfloat[][Napper]{
    \includegraphics[width=0.83\columnwidth]{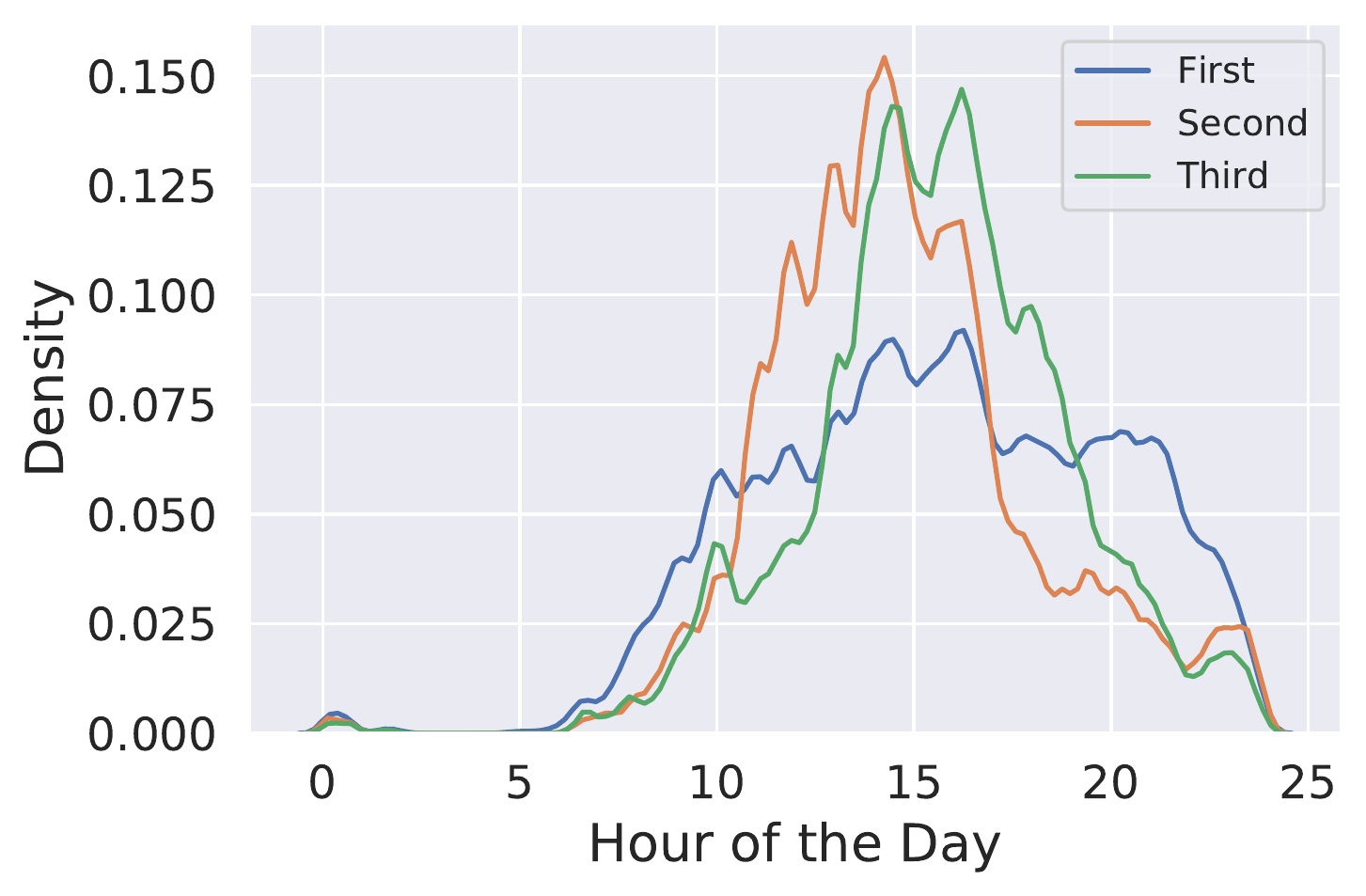}
    } \\
    \subfloat[][Afternoon]{
    \includegraphics[width=0.83\columnwidth]{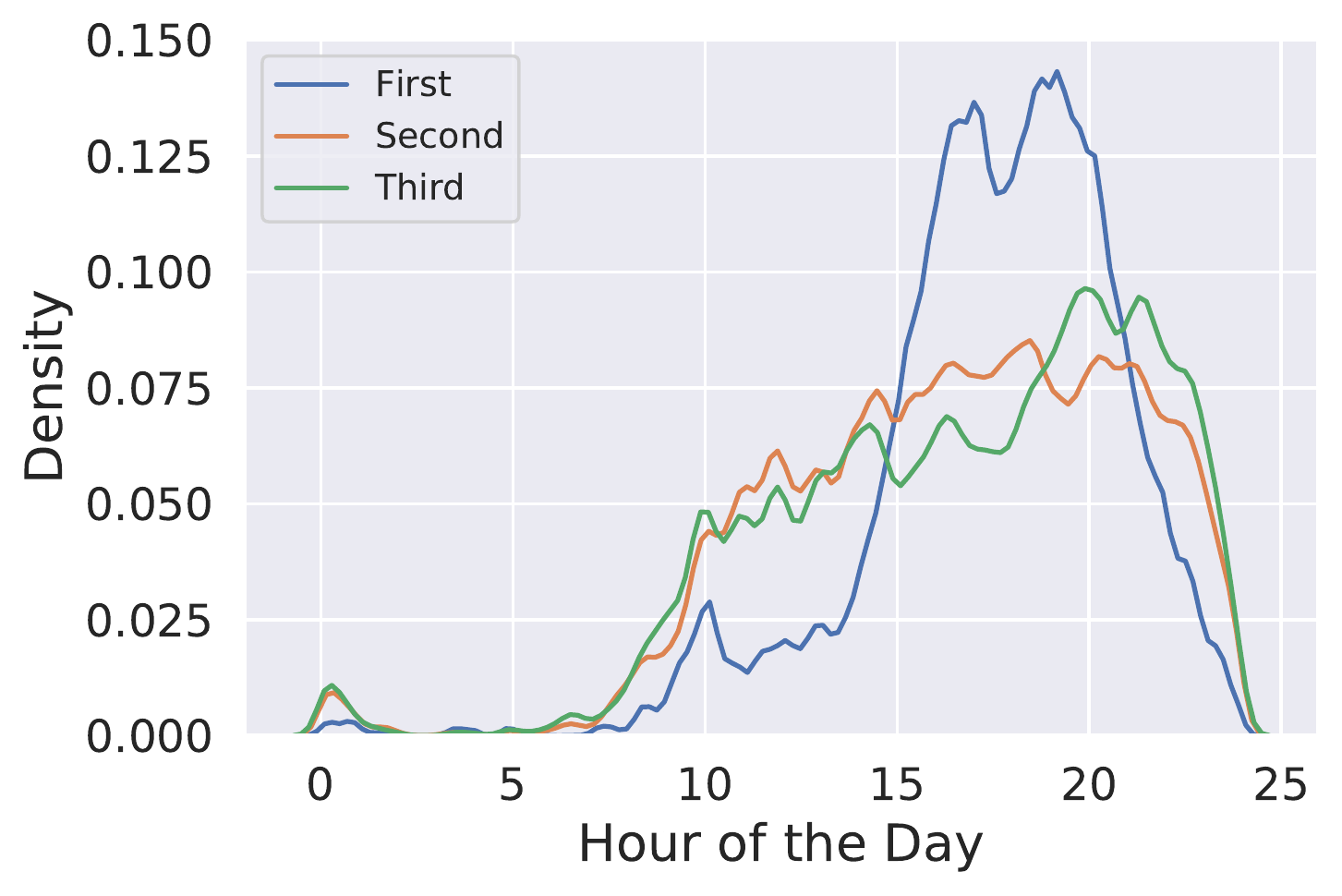}
    }
    \subfloat[][Evening]{
    \includegraphics[width=0.83\columnwidth]{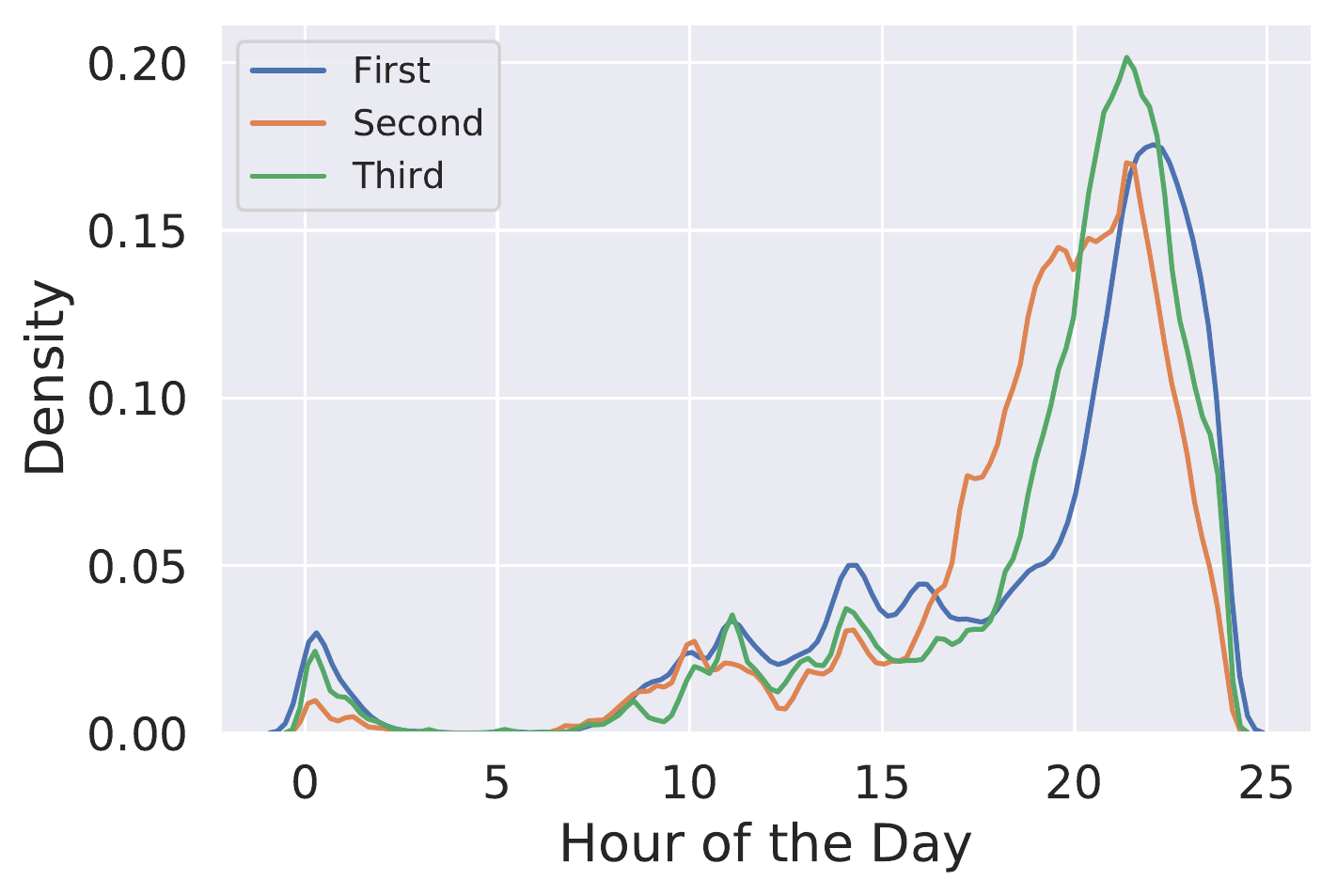}
    }
    \caption{Comparison of keystroke distributions for each proposed chronotype across feature vector definitions for the US dataset. Vector binning strategies are: First = [3-9, 9-15, 15-21, 21-3]; Second = [5-11, 11-17, 17-23, 23-5]; Third = [7-13, 13-19, 19-1, 1-7].}
    \label{fig:dists-across-feature-vectors}
\end{figure*}

\subsection{Comparison of binning strategies}
\label{sec:results-binning-robustness}

Fig.~\ref{fig:dists-across-feature-vectors} shows keystroke distributions for each proposed chronotype across feature vector definitions for the US context. In the clustering labeled "First" we described each student with the four-dimensional vector used in the remainder of our analysis, with percentage of keystrokes occurring in each of four 6-hour time intervals starting at 3:00am. The "Second" clustering used time intervals starting at 5:00am and the "Third" started at 7:00am.

\begin{table*}[]
    \centering
    \caption{Statistics of the clusters for the two contexts, US and European. The four \textit{Centroid} columns indicate the 4D centroid discovered in $k$-means clustering. \textit{Students} is the number of students in each cluster. \textit{Keystrokes} is the median number of keystrokes per student. \textit{Homework average score} is on project A5 (max of 100 points) for the US context and on all homework programming assignments (max of 215 points) for the European context. \textit{Exam average score} is on the first midterm (max of 100 points) for the US context and is a combined score on the two exams (max of 30 points combined) for the European context. \textit{Hours before deadline} is calculated by finding the median keystroke timestamp for each week and averaged over all weeks. Standard deviations are included with the homework/exam scores and hours before the deadline averages.}
    \begin{tabular}{lcccccccccc}
    \toprule
    &&\multicolumn{4}{c}{Centroid}&& &\multicolumn{2}{c}{Avg score} & Hours \\
    Context & Cluster & 3-9&9-15&15-21&21-3 & Students & Keystrokes & Homework & Exam & before deadline \\
    \midrule
    US & morning & 0.03 & 0.70 & 0.19 & 0.08 & 24\% (123) & 21912 & $84 \pm 28$ & $79 \pm 12$ & $35 \pm 32$ \\
    \rowcolor{table-highlight}
    European & morning & 0.02 & 0.67 & 0.27 & 0.04 & 11\% (36) & 58992 & $162 \pm 56$ & $22 \pm 11$ & $85 \pm 36$ \\
    US & napper & 0.06 & 0.39 & 0.43 & 0.12 & 40\% (208) & 25078 & $83 \pm 29$ & $77 \pm 12$ & $33 \pm 36$ \\
    \rowcolor{table-highlight}
    European & napper & 0.01 & 0.42 & 0.44 & 0.12 & 31\% (98) & 70949 &  $172 \pm 47$&  $24 \pm 8$ & $59 \pm 34$ \\
    US & afternoon & 0.01 & 0.12 & 0.76 & 0.11 & 21\% (111) & 18008 & $73 \pm 35$ & $74 \pm 14$ & $21 \pm 30$ \\
    \rowcolor{table-highlight}
    European & afternoon & 0.01 & 0.19 & 0.60 & 0.20 & 29\% (91) & 60427 & $165 \pm 48 $&  $25 \pm 9$  & $ 46 \pm 38 $\\
    US & evening & 0.01 & 0.16 & 0.29 & 0.53 & 15\% (77) & 17970 & $70 \pm 36$ & $71 \pm 15$ & $14 \pm 29$ \\
    \rowcolor{table-highlight}
    European & evening & 0.03 & 0.15 & 0.35 & 0.46 & 29\% (93) & 53595 & $159 \pm 57$&   $22 \pm 11$ & $46 \pm 37$ \\
\bottomrule
    \end{tabular}
    \label{tab:clusters}
\end{table*}

\subsection{Chronotypes learned from clustering}
\label{sec:results-clustering}
Our first research question is: \textit{What chronotypes would be discovered from clustering keystroke data collected from two contexts?} Related questions include whether the chronotypes between the two contexts match each other and whether they match those found in the literature.

\begin{figure*}
\centering
\begin{tabular}{m{1mm}c|c}
    &\textbf{Morning} & \textbf{Napper} \\
    \raisebox{5\normalbaselineskip}[0pt][0pt]{\rotatebox{90}{\textbf{US}}} & \includegraphics[width=0.73\columnwidth]{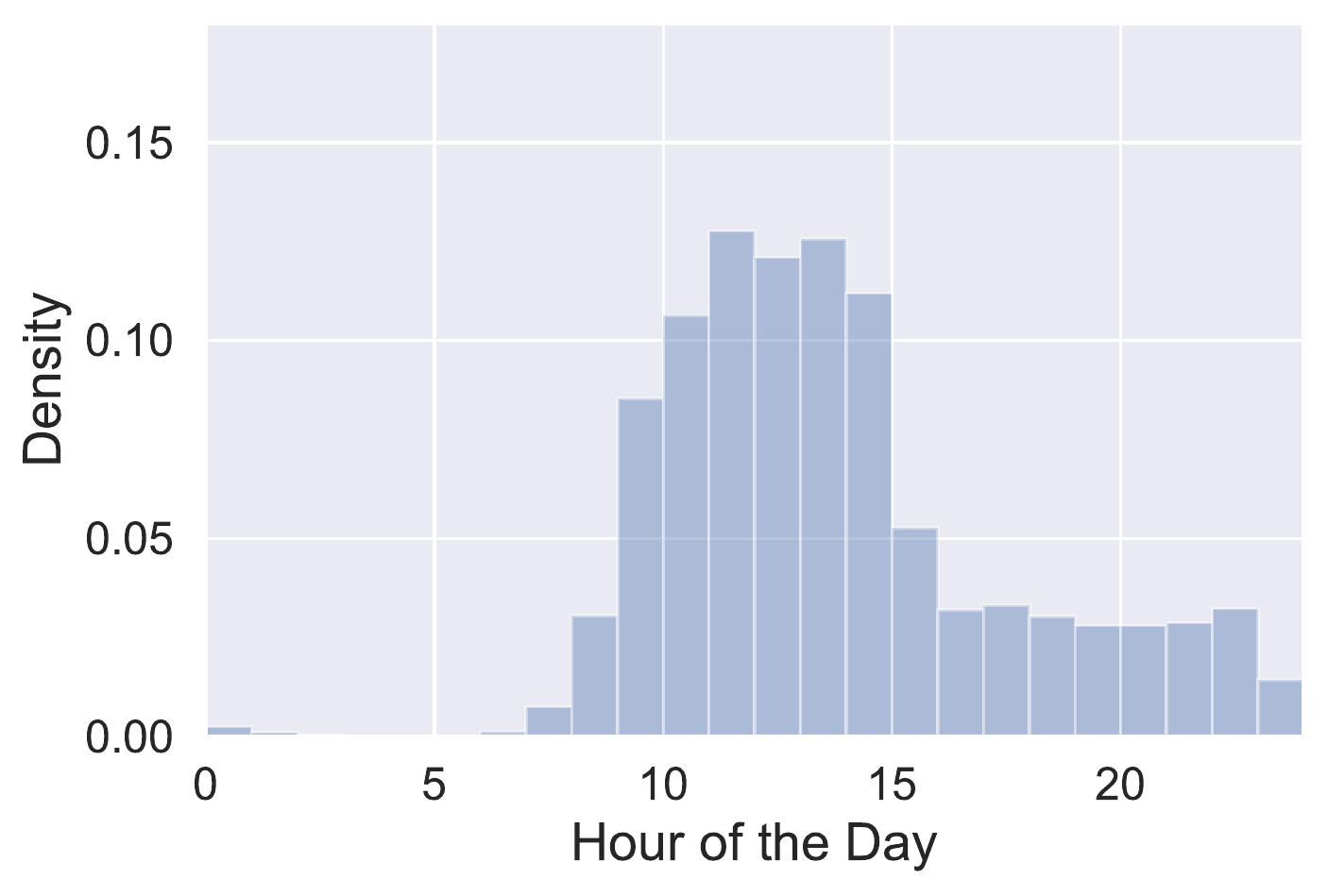} &
    \includegraphics[width=0.73\columnwidth]{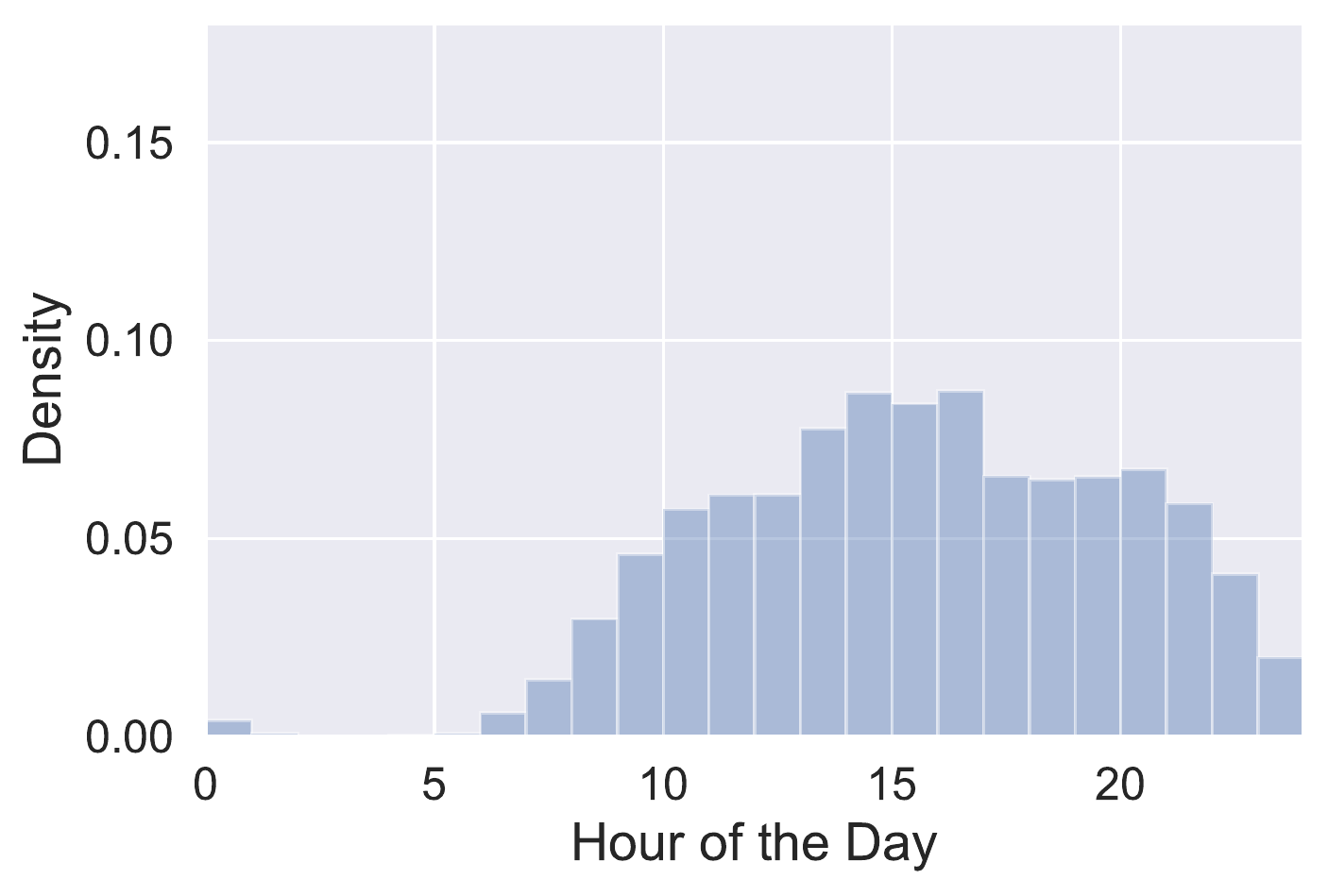} \\
    \raisebox{4\normalbaselineskip}[0pt][0pt]{\rotatebox{90}{\textbf{European}}} & \includegraphics[width=0.73\columnwidth]{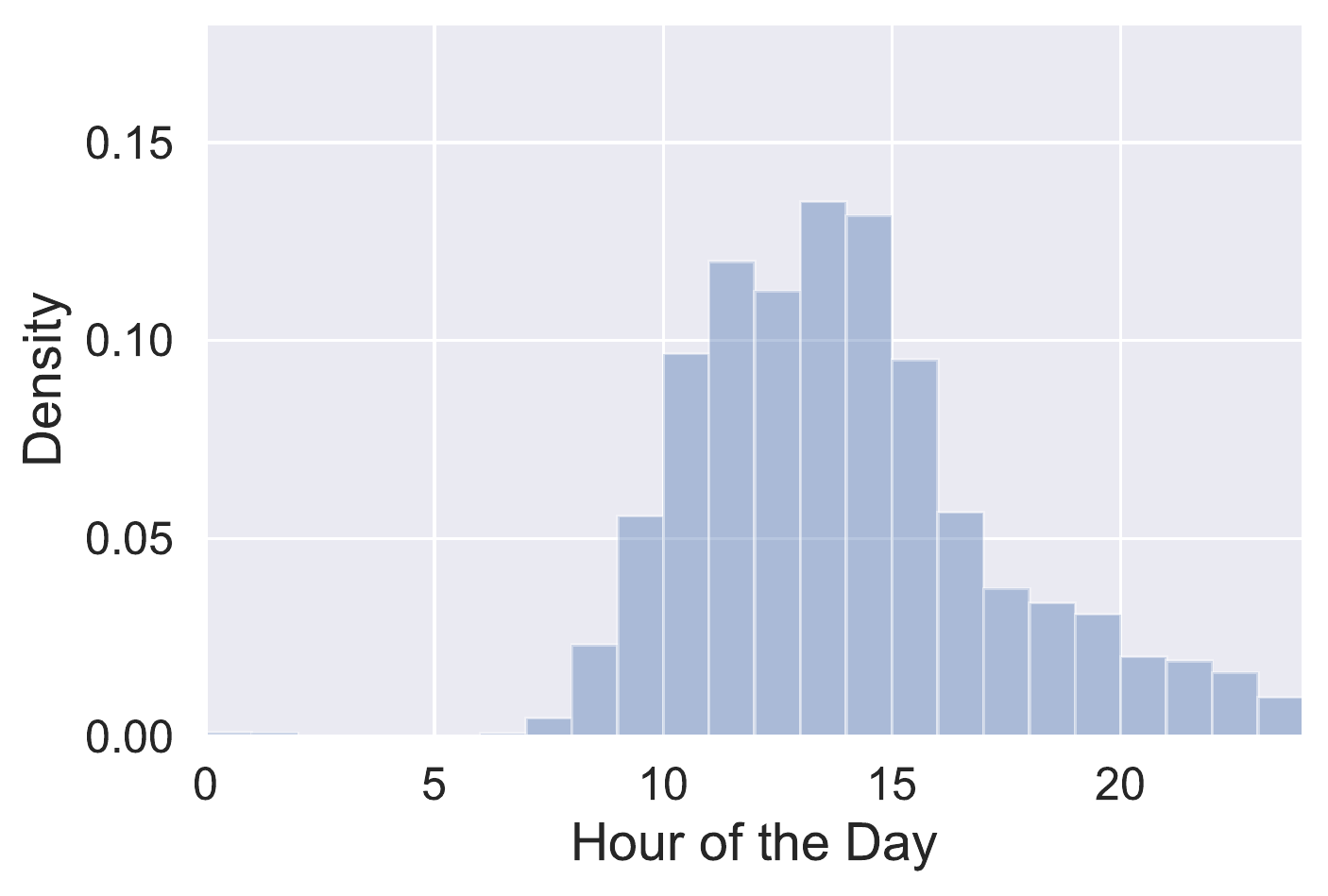} &
    \includegraphics[width=0.73\columnwidth]{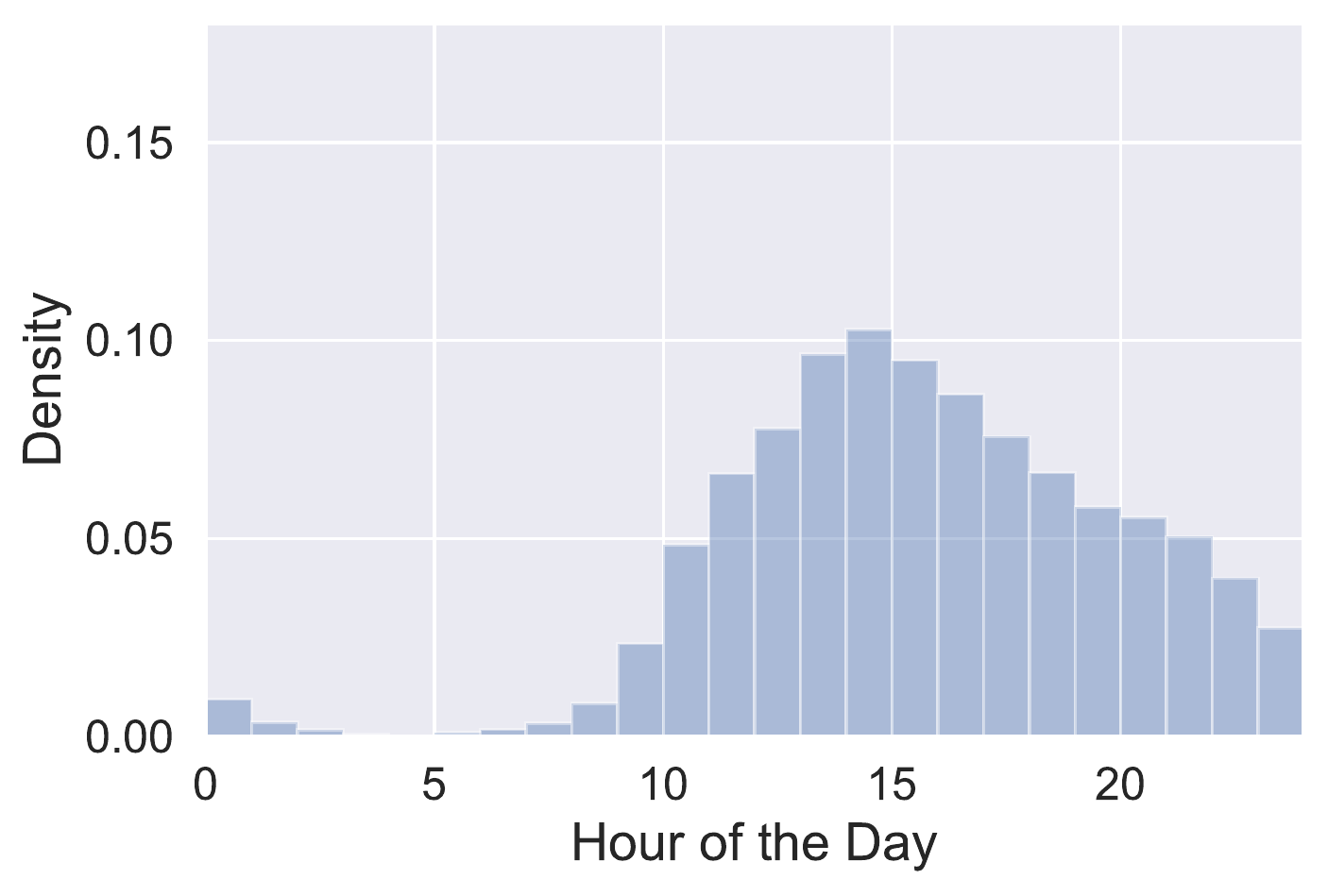} \\
     \midrule
    &\textbf{Afternoon} & \textbf{Evening} \\
    \raisebox{5\normalbaselineskip}[0pt][0pt]{\rotatebox{90}{\textbf{US}}} & \includegraphics[width=0.73\columnwidth]{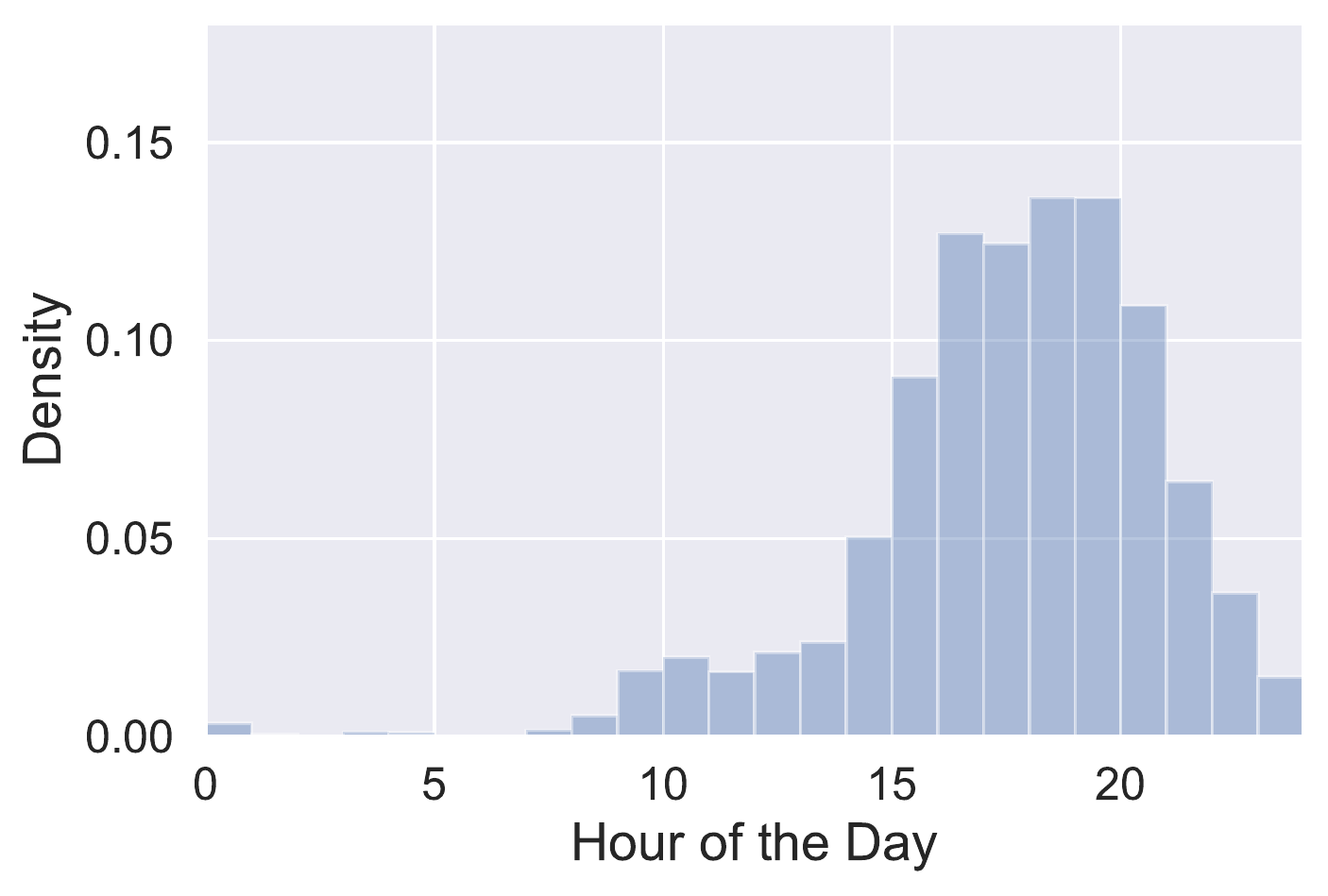} &
    \includegraphics[width=0.73\columnwidth]{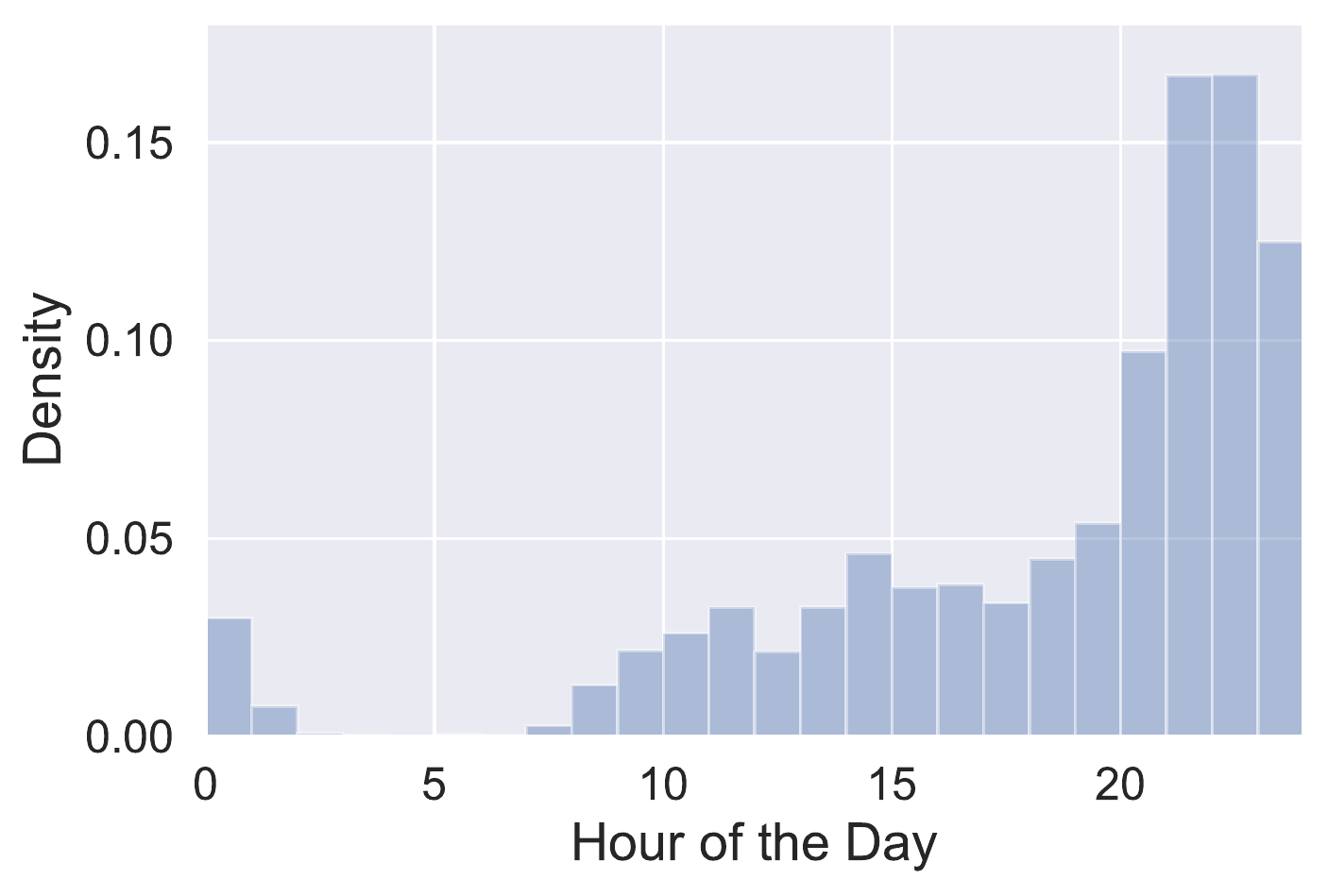} \\
    \raisebox{4\normalbaselineskip}[0pt][0pt]{\rotatebox{90}{\textbf{European}}} & \includegraphics[width=0.73\columnwidth]{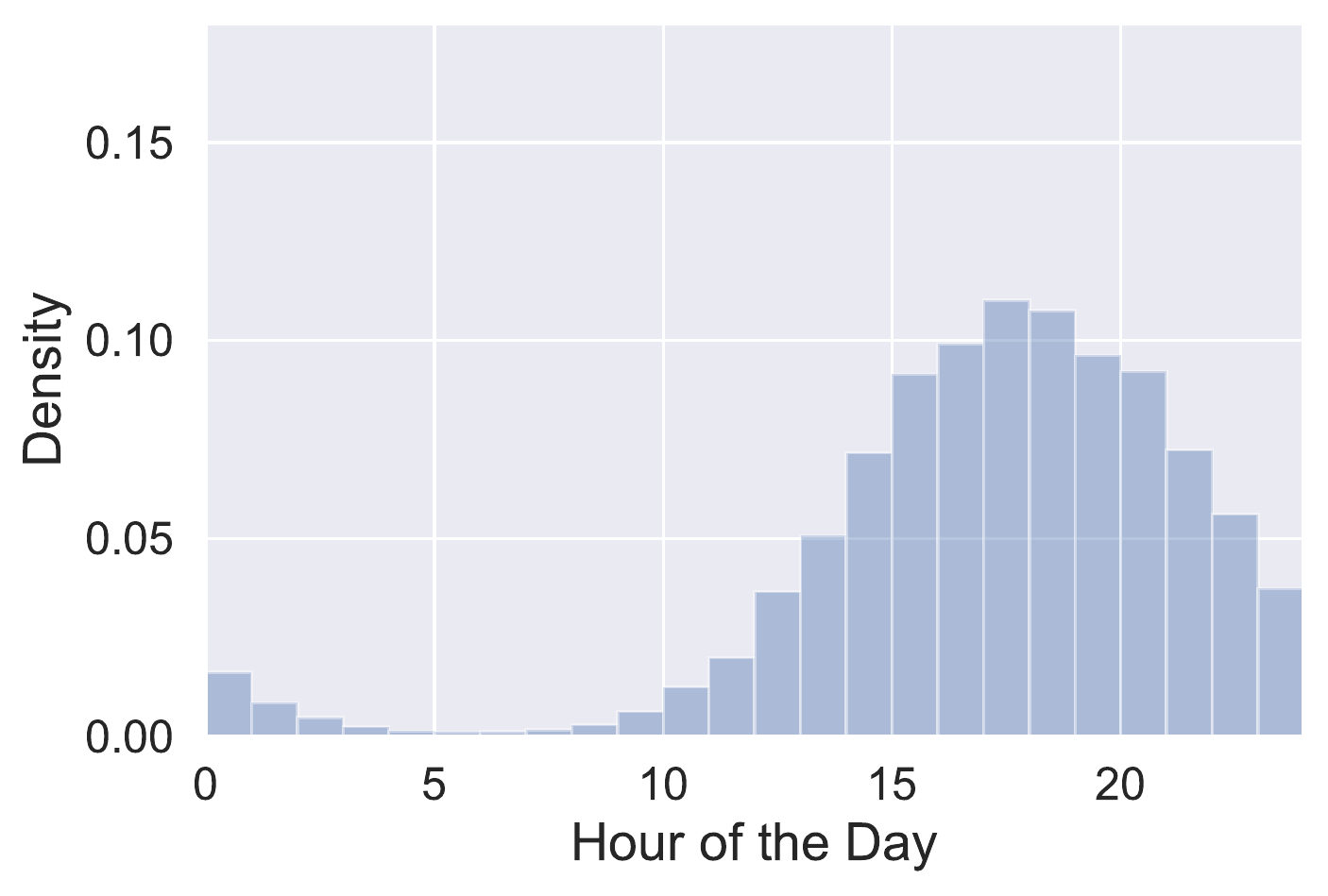} &
    \includegraphics[width=0.73\columnwidth]{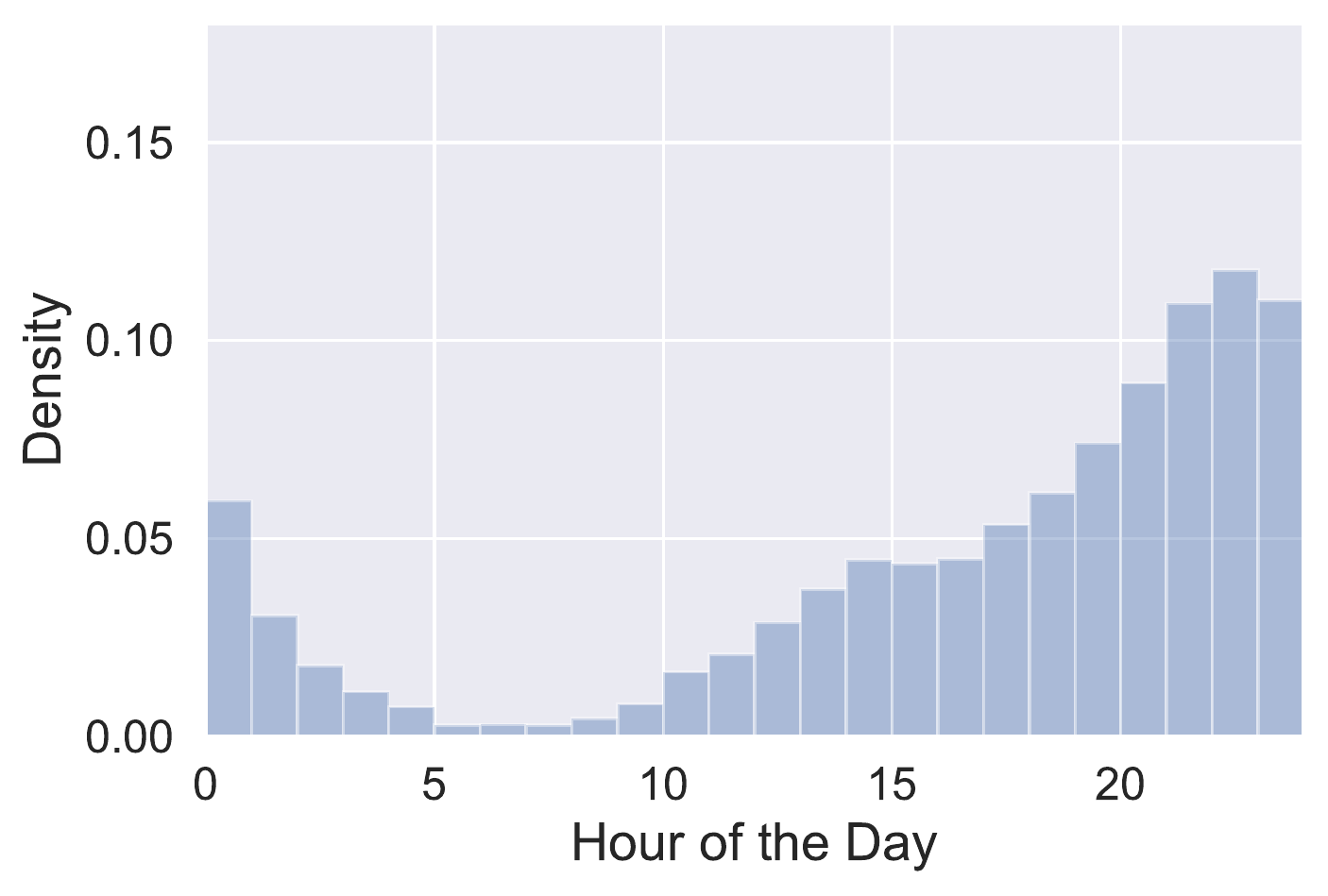}
\end{tabular}
    \caption{Distribution of keystrokes for the four clusters for each context labeled with proposed chronotypes.} 
    \label{fig:dists-all}
\end{figure*}

Fig.~\ref{fig:dists-all} shows keystroke distributions for each of the four clusters for each of the two contexts. It is notable how similar the four clusters are between the two contexts. Considering the differences in context, including country, latitude (over $18\deg$ difference), average daylight (roughly 3 hours), instructor, course structure, homework composition, course pacing, and class time, the distributions are remarkably similar. Indeed, hours remaining before due date and number of keystrokes (Fig.~\ref{fig:outcomes}) show similar behaviors across clusters even given the context differences.

Due to distinctive differences between the discovered clusters and their high similarity to the Putilov's et al. \cite{Putilov2019} chronotypes, we assigned the same terms to the clusters (Fig.~\ref{fig:dists-all}). For the morning cluster, peak activity is from 10am to 3pm with activity trailing off into the evening. The napper cluster has the widest range of active times. The afternoon cluster has peak activity from 1pm to 5pm, but activity is also reasonably high from 10am to 1pm. The evening cluster starts around 9am with peak activity from 8pm to 11pm.

\subsection{Chronotype and relation to course outcomes}
In this section we explore the following research question: \textit{How do the typical working times of students relate to academic outcomes?} In the following discussion, refer to Table~\ref{tab:significance} for results of Kruskal-Wallis H significance testing between clusters. We first look at correlations of chronotype with project and exam scores. In Fig.~\ref{fig:outcomes} (\textit{Projects}) we see a pronounced difference in median project score between the four chronotypes in the US context but much less so in the European context, and indeed, there are strongly significant differences in the US context and no statistically detectable differences in the European context (Table~\ref{tab:significance}). We see similar phenomena relating to exam scores in Fig.~\ref{fig:outcomes} (\textit{Exams}): for the US context, strongly significant differences exist whereas the European context had almost no difference, although the European context may have been influenced by the ceiling effect in this case. For both project and exam scores in the US context, morning and napper students performed better than afternoon and evening students.

We also considered the correlation between chronotype and how long before the due date students work on their assignments (\textit{Hours remaining} in Fig.~\ref{fig:outcomes}). Both contexts showed strongly significant differences, with students who work on their assignments in the morning generally working earlier relative to the due date than students who work in the evening and night time. Similarly, Fig.~\ref{fig:outcomes} (\textit{\# keystrokes}) shows the number of keystrokes executed by students. Students in the evening chronotype typed fewer characters, on average, than students in the other chronotypes. We discuss implications of these results in Section~\ref{sec:discussion}.

\begin{figure*}
\centering
\begin{tabular}{m{1mm}cc}
    &\textbf{US} & \textbf{European} \\
    \raisebox{4.5\normalbaselineskip}[0pt][0pt]{\rotatebox{90}{\textbf{Projects}}} & \includegraphics[width=0.73\columnwidth]{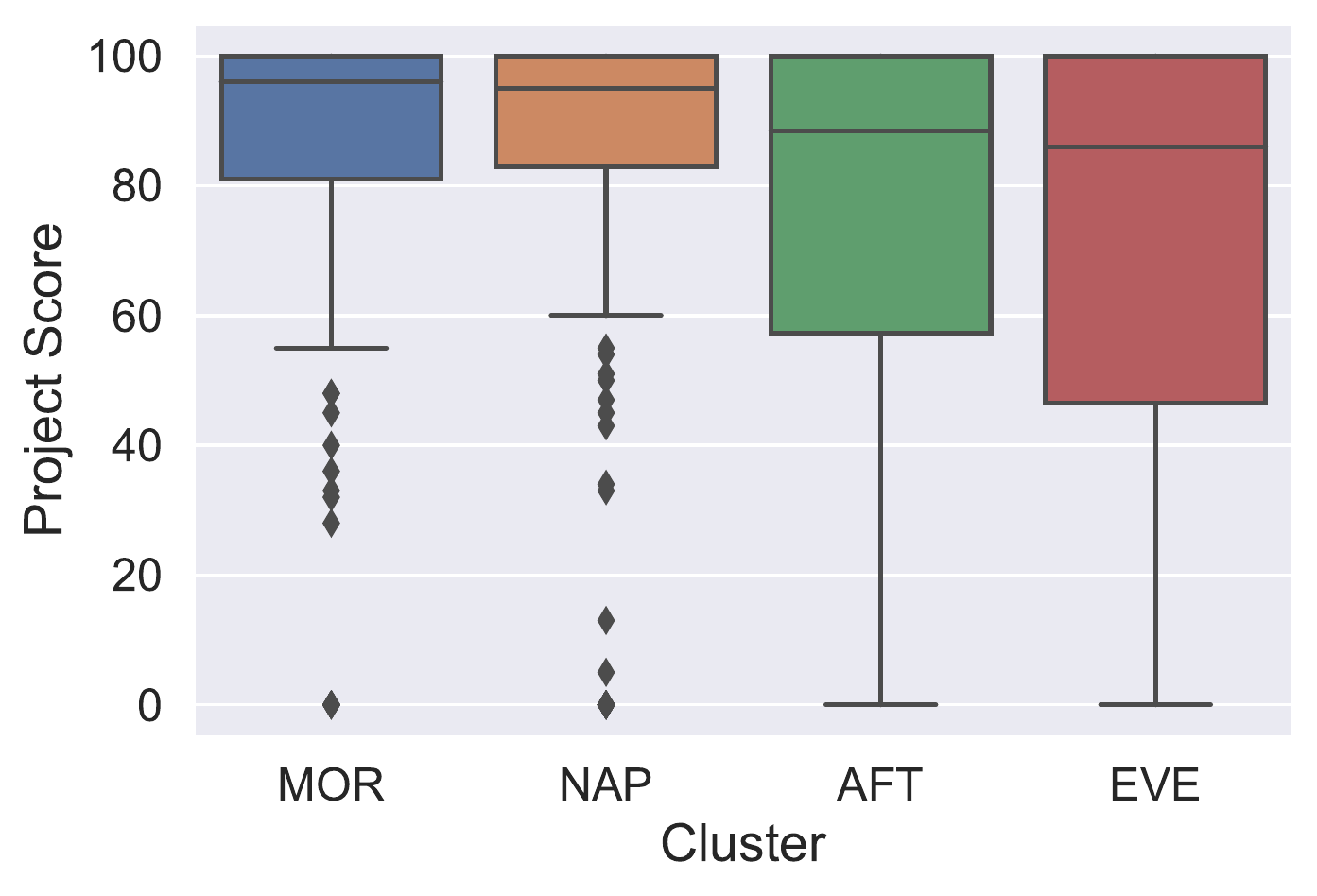} &
    \includegraphics[width=0.73\columnwidth]{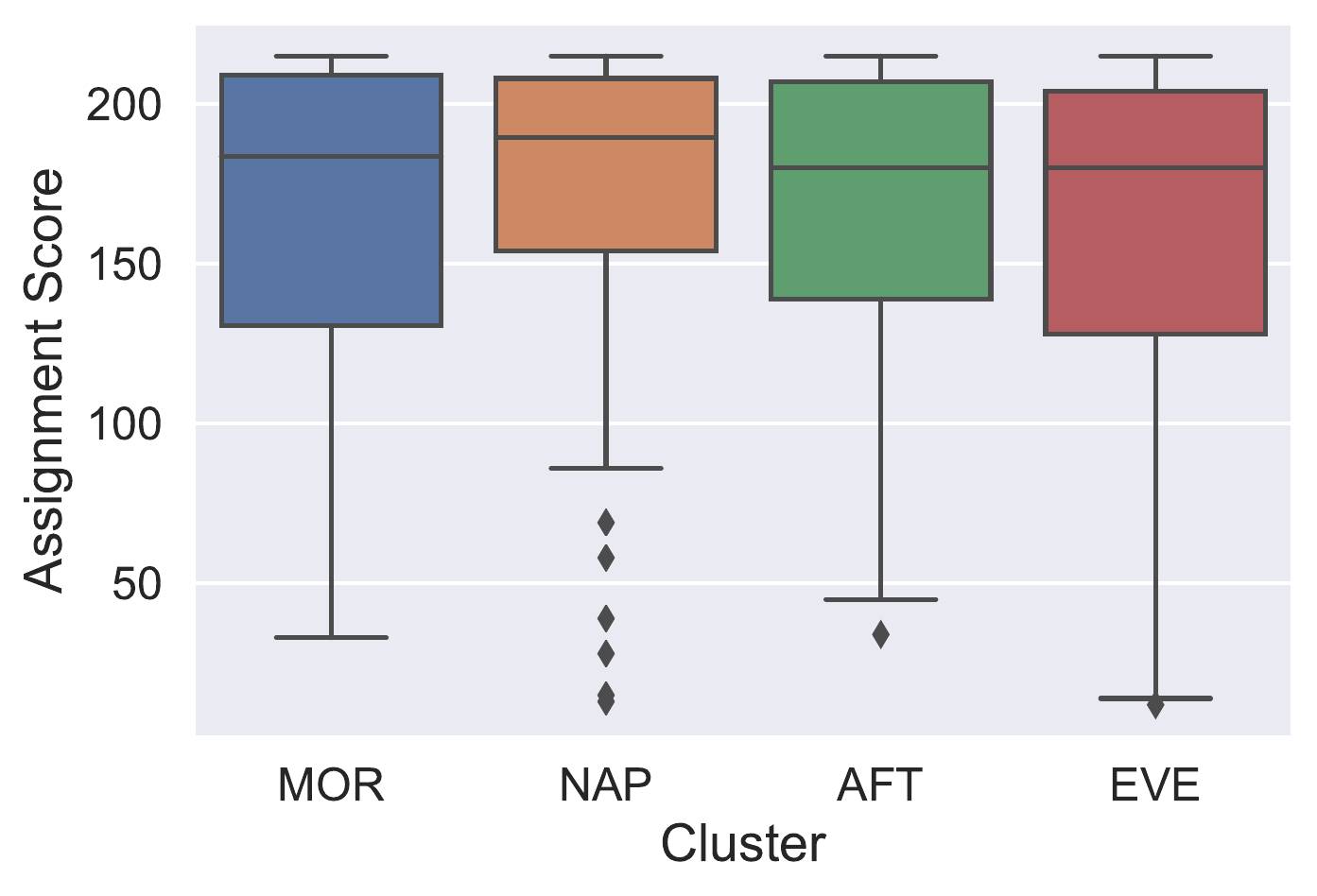} \\
    \raisebox{4.5\normalbaselineskip}[0pt][0pt]{\rotatebox{90}{\textbf{Exams}}} & \includegraphics[width=0.73\columnwidth]{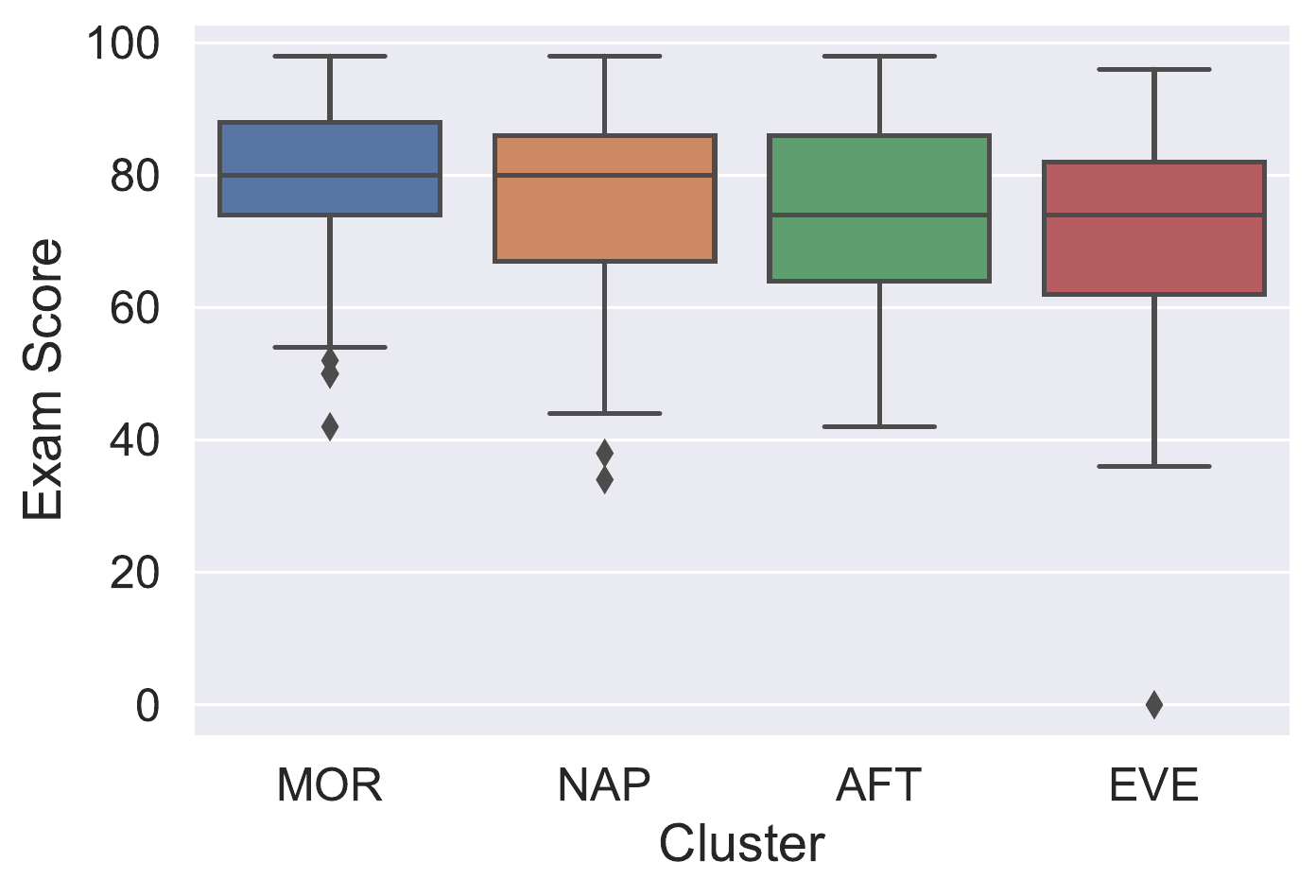} &
    \includegraphics[width=0.73\columnwidth]{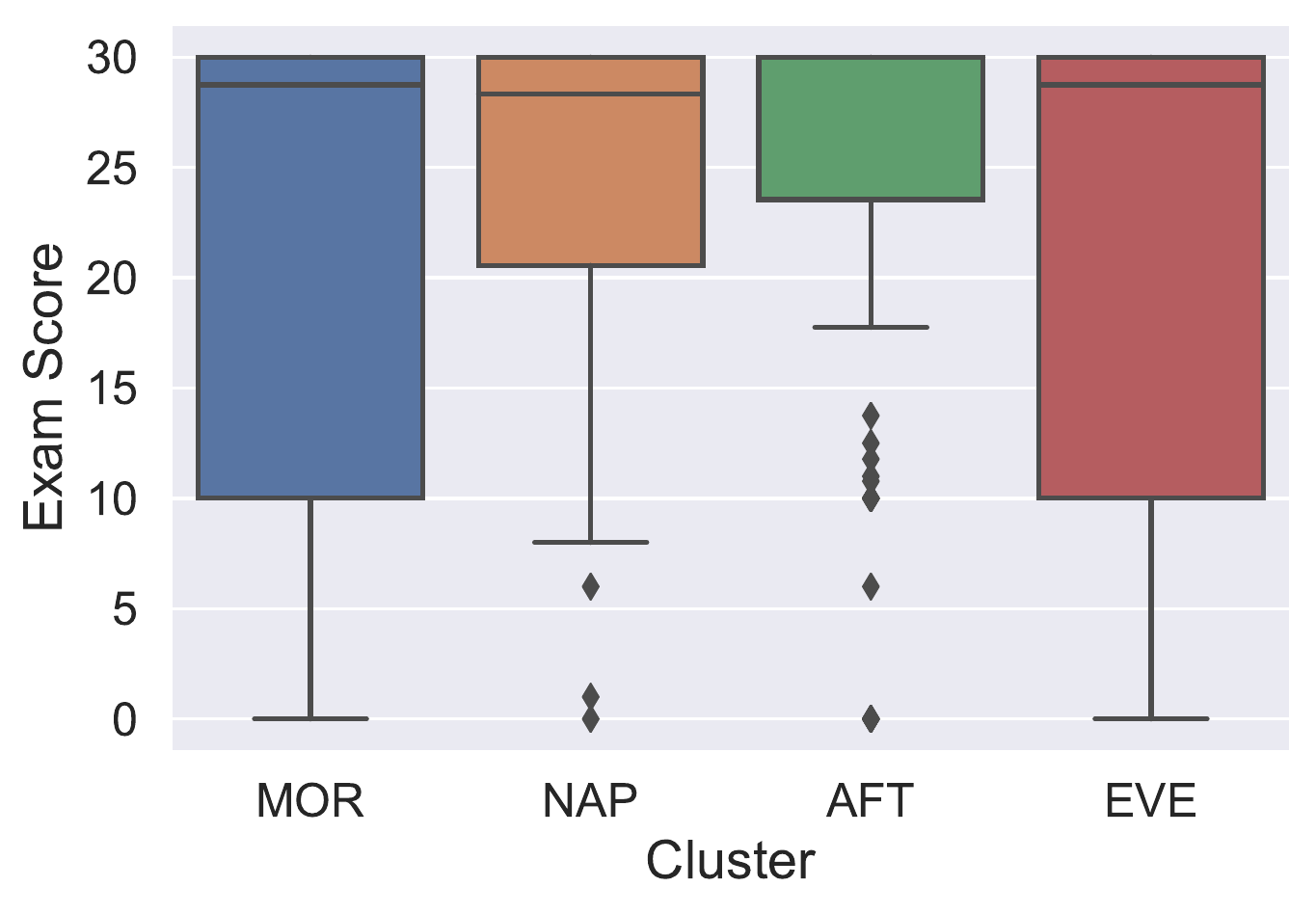} \\
    \raisebox{3\normalbaselineskip}[0pt][0pt]{\rotatebox{90}{\textbf{Hours remaining}}} & \includegraphics[width=0.73\columnwidth]{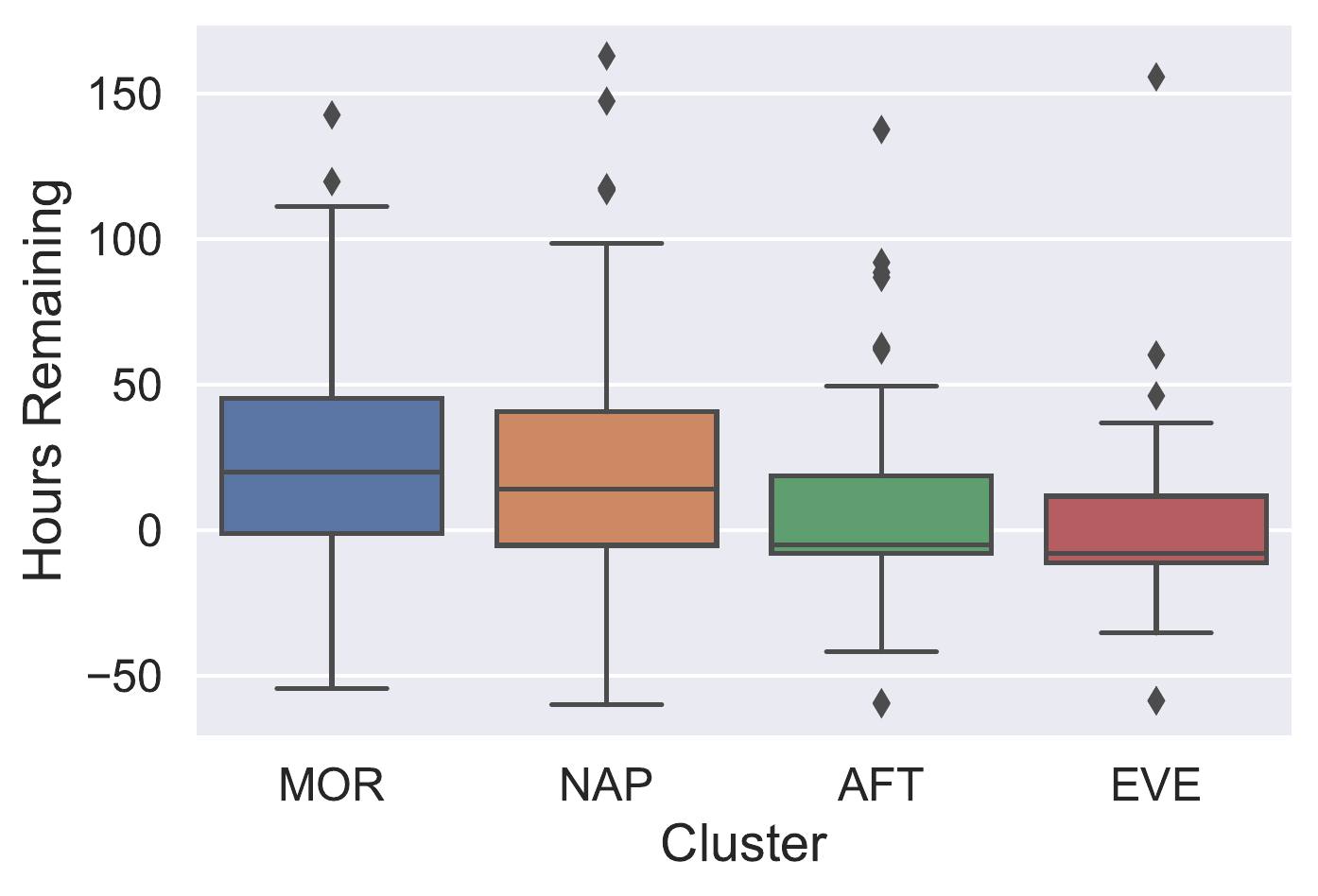} &
    \includegraphics[width=0.73\columnwidth]{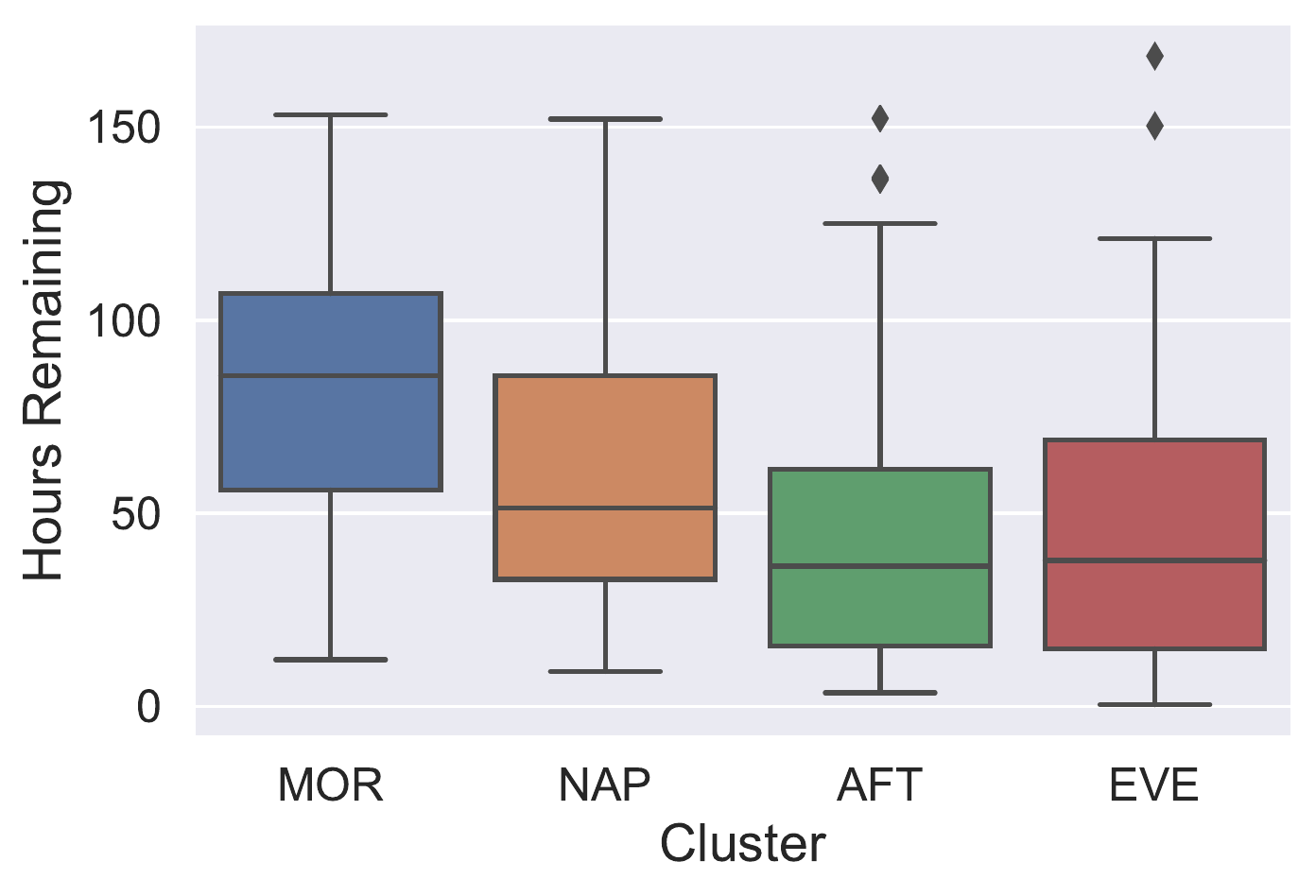} \\
    \raisebox{3.4\normalbaselineskip}[0pt][0pt]{\rotatebox{90}{\textbf{\# keystrokes}}} & \includegraphics[width=0.73\columnwidth]{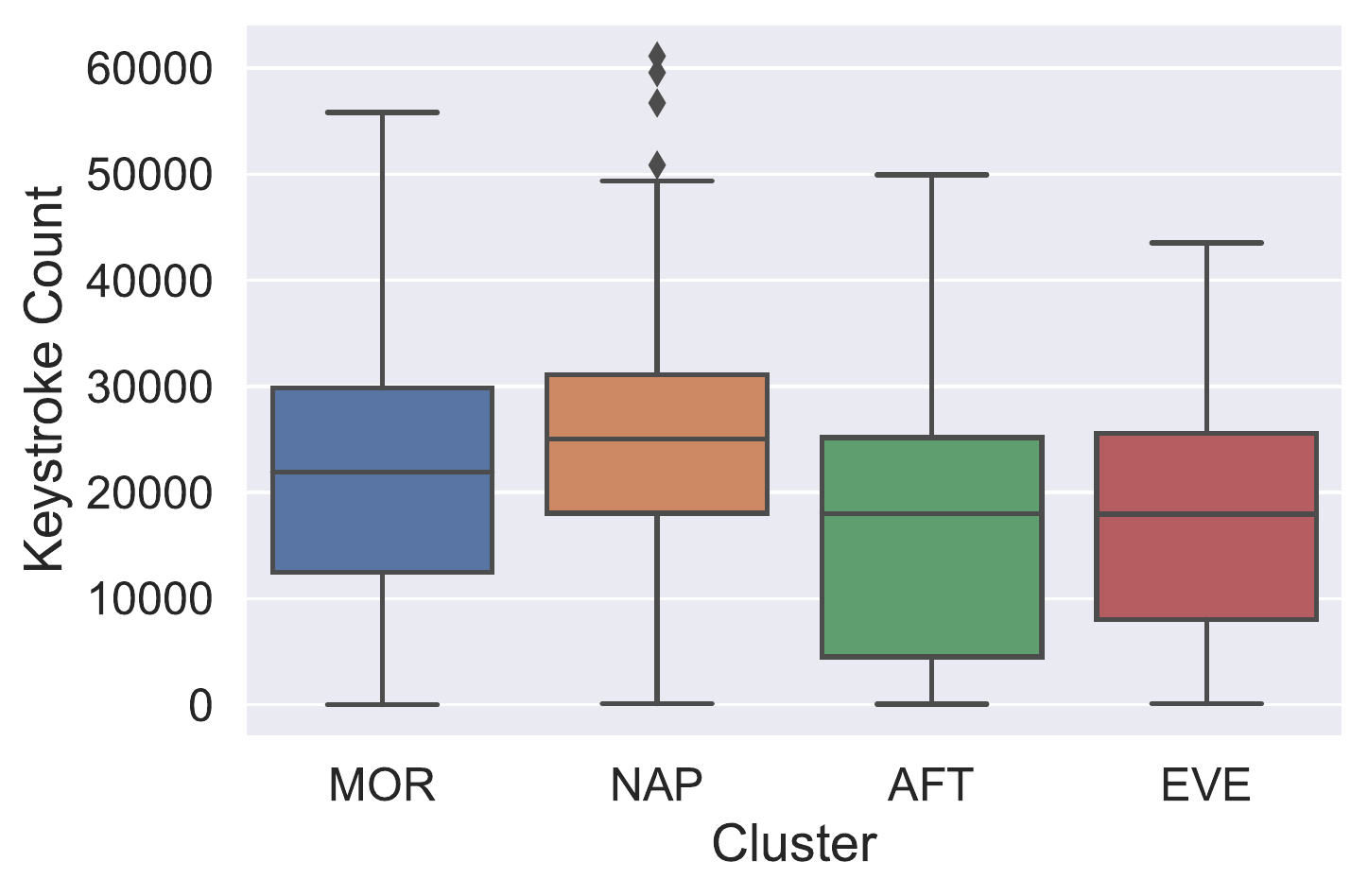} &
    \includegraphics[width=0.73\columnwidth]{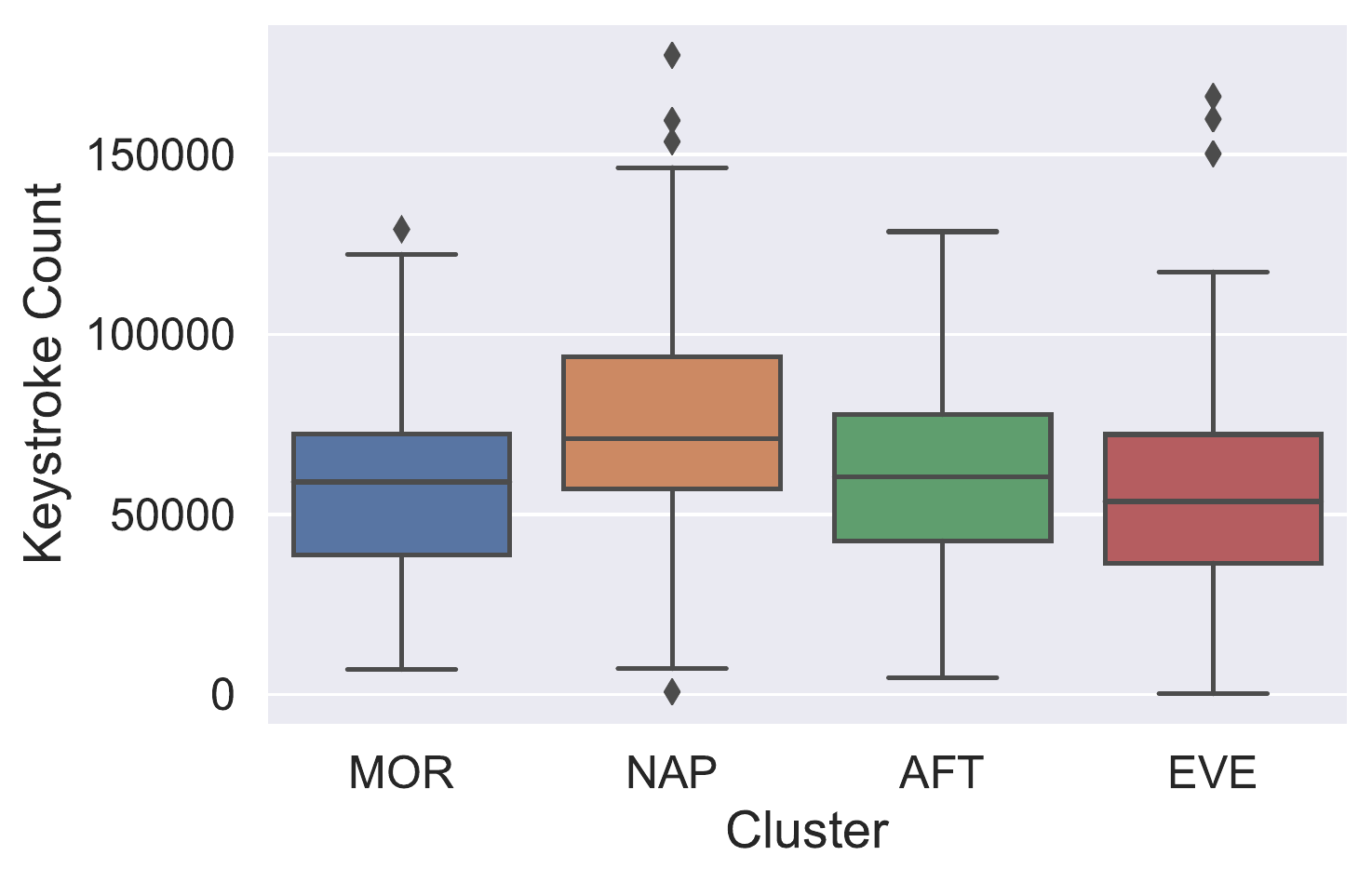} \\
\end{tabular}
    \caption{Outcomes by cluster: project score, exam score, hours remaining before due date, and number of keystrokes. Whiskers extend 1.5$\cdot$IQR past the low and high quartiles.
    }
    \label{fig:outcomes}
\end{figure*}

\subsection{Differences in context}
\label{sec:context-diffs}
In this section we report results of the research question: \textit{How do contextual factors affect academic outcomes?} \red{We first note from Table~\ref{tab:clusters} that, while the centroids of the clusters are remarkably similar between the US and European contexts, the distribution of students among the clusters is different. The majority of US students have morning and napper chronotypes, whereas most European students are afternoon and evening. A number of factors could be causing this difference, including culture, demographic, and hours of sunlight per day (Table~\ref{tab:ctx-summary}). Further investigation is needed to understand the cause of the differences in distribution.}

One of the most notable differences between the US and the European contexts is the time before the due date of the keystrokes. See Fig.~\ref{fig:outcomes} (\textit{Hours remaining}). For two of the US clusters, the median keystroke is actually \textit{later} than the due date (note from Table~\ref{tab:clusters} that the mean, however, is before the due date), while the median keystrokes for all four clusters in the European context are well before the due date. From Table~\ref{tab:clusters} we see that, in every chronotype, students in the European context work on their projects at least 25 hours earlier, on average, than students in the US context, and European morning students work fully 50 hours earlier on average. We propose two contextual differences contributing to this effect. The first is the late-work policy. In the US context students could turn in late work for reduced points. In the European context, late work was not accepted. The second factor is project composition. It has been shown that when projects are broken into multiple smaller pieces, students tend to start earlier~\cite{denny2018improving}. Indeed, our results support this conclusion: in the US context the projects are broken into two sub-projects, while the European context has tens of sub-projects due each week. Considering the factors of late-work policy and multiple, smaller projects, it isn't surprising that a difference in time before the due date exists between the contexts. What is perhaps surprising, however, is the magnitude of the difference.

\begin{table}[]
    \centering
    \caption{Results of running a Kruskal-Wallis H significance test across clusters for different outcomes.}
    \begin{tabular}{lcccc}
    \toprule
        & \multicolumn{2}{c}{US} & \multicolumn{2}{c}{European} \\
        & $H$ & $p$ & $H$ & $p$ \\
    \midrule
        Assignment scores & $13.178$ & $0.004$ & $2.386$ & $0.496$ \\
        Exam & $18.894$ & $<0.001$ & $4.657$ & $0.199$ \\
        Hours remaining & $58.162$ & $<0.001$ & $35.667$ & $<0.001$ \\
        \# keystrokes & $40.947$ & $<0.001$ & $21.054$ & $<0.001$ \\
    \bottomrule
    \end{tabular}
    \label{tab:significance}
\end{table}

\section{Discussion \label{sec:discussion}}

\subsection{Robustness of clustering}

In order to avoid the fragility of clustering in higher dimensions, we clustered four-dimensional feature vectors into chronotypes. To mitigate the risk of biasing the clusters by quantization strategy we clustered two additional times with different bins. Fig.~\ref{fig:dists-across-feature-vectors}, which shows distributions of keystrokes for each proposed chronotype across feature vector definitions, indicates that clusters of students are reasonably robust to binning strategy.

\subsection{Robustness to external factors}
\label{sec:external-factors}

Many factors beyond chronotype have the potential to influence how students use their time, including other courses, jobs, family obligations, social engagements, and recreation. The extent to which these external factors influence when students work on their programming assignments and whether discovered clusters indeed represent students' chronotypes are two important questions. 
For example, we claim that students in our \emph{evening} group are of the evening chronotype, but one could claim that they are actually students with day jobs and are thus constrained to complete their programming assignments in the evenings.

Looking separately at working days and weekends (Fig.~\ref{fig:java-keystrokes-weekday-weekend}) -- two time periods that generally have little in common with respect to jobs, class times, and other responsibilities -- we see that external factors have little effect on when students work on their assignments: patterns of activity times remain reasonably constant. Using our previous example, evening students tended to work on their assignments in the evening whether it was during the week or on the weekend, suggesting that even though they may have the option to work on homework in the morning during weekends, they still choose to do so in the evening, an indication that times students work on their assignments are governed primarily by diurnal preference.

To test the similarity of the weekday and weekend distributions in Fig.~\ref{fig:java-keystrokes-weekday-weekend} we ran Mann-Whitney U tests on the distributions for all chronotypes (only the evening chronotype is shown in the figure) shifted by 7 hours to make the distributions unimodal. Because the cyclic nature of the data could compromise the integrity of a rank-based test, we also ran chi-squared tests using the 24 one-hour bins shown in the figures. Both tests for each of the four chronotypes yielded $p<0.0001$. However, all effect sizes (Cohen's $d$ for Mann-Whitney U and Cramer's $V$ for chi-squared) were 0.2 or less, so, while behavior is different between weekday and weekend, the difference is small.

\begin{figure*}
    \centering
    \subfloat[][Weekday]{
        \includegraphics[width=0.7\columnwidth]
        {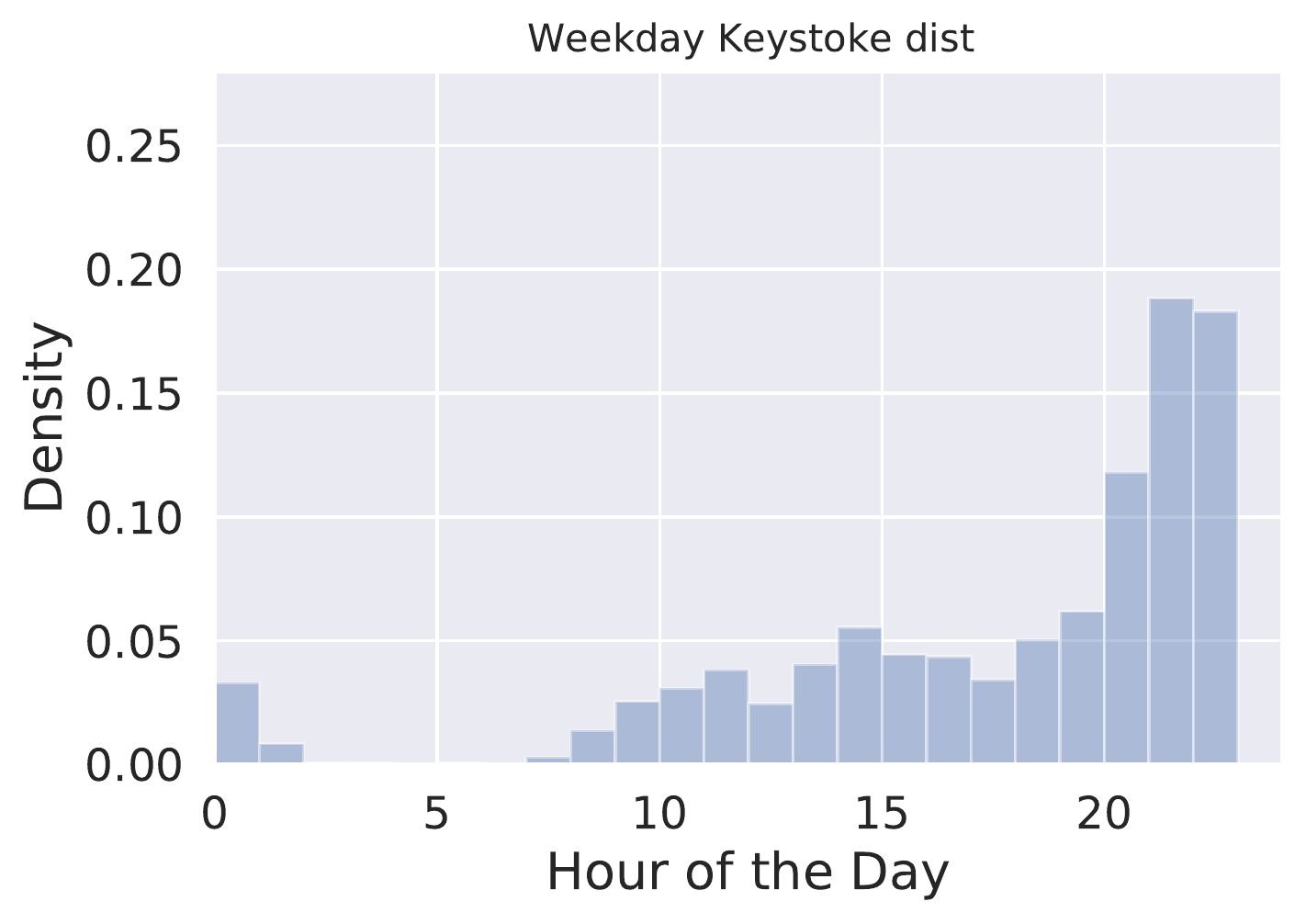}
        \label{fig:java-keystrokes-evening-weekday}
    }    
    \subfloat[][Weekend]{
        \includegraphics[width=0.7\columnwidth]
        {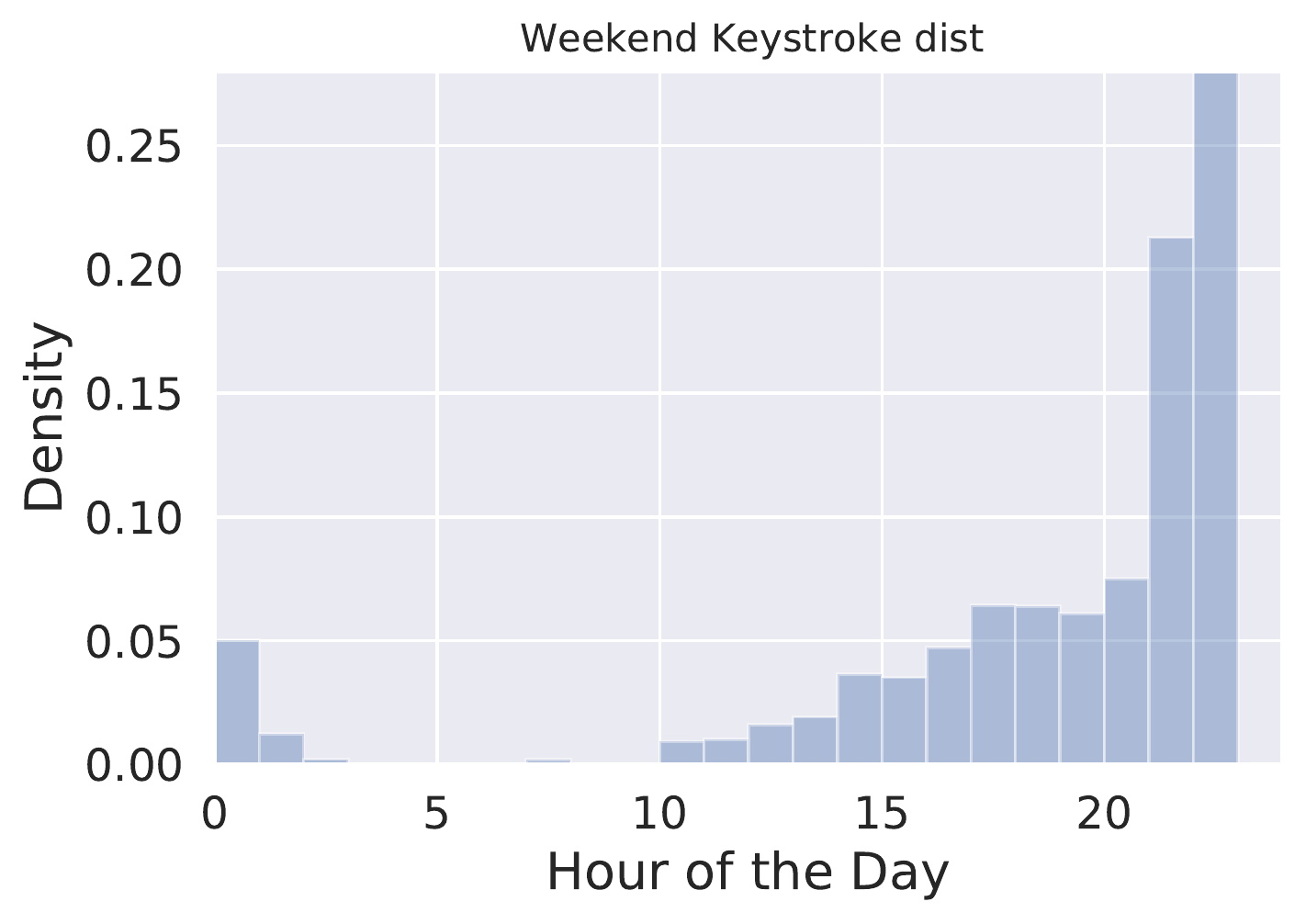}
        \label{fig:java-keystrokes-evening-weekend}
    } \\
    \caption{Keystroke distribution of the evening chronotype in the US split into weekday and weekend keystrokes.}
    \label{fig:java-keystrokes-weekday-weekend}
\end{figure*}

\subsection{General Insights}

There is a common stereotype of computer programmers as being night owls. A striking result of our data is that, at least among CS1 students, this stereotype does not hold up. The majority of keystrokes in our study were done between 9am and 5pm (54\%) and most of the remaining keystrokes were executed between 5pm and 9pm (27\%). Only 1\% of keystrokes occurred between midnight and 5:00am.
Our results support prior work that has found that the majority of programmers follow typical working hours~\cite{Claes2018}. Indeed, even among the few students in the evening chronotype, most keystrokes were logged before midnight.

Another important result is that, while context appeared to have a strong effect on correlations between chronotype and outcomes, the chronotypes themselves found through unsupervised learning were very similar between the two contexts.
The fact that similar chronotypes were found, both in terms of cluster centroids (Table~\ref{tab:clusters}) and keystroke distributions (Fig.~\ref{fig:dists-all}) between two very different contexts is a strong support for chronotype theory and its characterizations~\cite{Putilov2019}.

\subsection{Effect on course outcomes}
A result exhibited in both the US and European contexts is that morning and napper students started working on weekly assignments much earlier than afternoon and evening students and nappers typed more characters in working on their assignments. Our results also provide some support for correlations between chronotype and academic achievement~\cite{preckel2011chronotype}: in the US context the evening chronotype was associated with lower assignment and exam scores when compared to the morning chronotype. However, the European context didn't show a difference. It is possible that the ceiling effect caused a lack of difference in project and exam scores between chronotypes in the European context. It is also possible that the design of programming assignments in the European context, as discussed in Section~\ref{sec:context-diffs}, could have affected academic outcomes across chronotypes, suggesting investigation into assignments with smaller pieces and fixed due dates.

\subsection{Synchrony with chronotype}

Research has found that scheduling learning activities during students' preferred working times increases academic achievement~\cite{dunn1987research,dunn1988learning}. Our data shows afternoon and evening students performing, in general, at a lower level. From the synchrony effect theory we consider the possibility that the university class schedule forced these students to work at suboptimal times, contributing to their sub-par behavior. To see if the imposed schedule affected their natural working time we compare each chronotype's keystroke distributions split between weekdays and weekends (the weekday/weekend distributions for the evening chronotype in the US context are shown in Fig.~\ref{fig:java-keystrokes-weekday-weekend}), as students would be expected to be more free to work according to their preferred schedule on weekends. As can be seen in the figure, students behaved very similarly on the weekends as they did during the week. Because of the small effect size (0.2, as reported above in Section~\ref{sec:external-factors}), which suggests that students worked similar hours whether during the week or on weekends, it appears that synchrony had only minimal effect on afternoon and evening student underachievement.

\subsection{Limitations}
We did not collect demographic or background information about study participants, which means that students who work at jobs or have other classes may have affected our chronotype distributions. However, the effect of these external factors appears to be minimal (see Section~\ref{sec:external-factors}).
Furthermore, \citet{Putilov2019} indicated that there is a group of people that doesn't belong to any of the four clusters and these classification-defying subjects may have diluted clarity of our clusters.

\section{Conclusion \label{sec:conclusions}}

In this article, we analyzed evidence for the existence of chronotypes using keystroke data collected from introductory programming students. To summarize, our research questions and their answers are as follows.

(RQ1) ``What chronotypes would be discovered from clustering keystroke data collected from two contexts?'' We identified four chronotypes similar to those discussed in the literature~\cite{Putilov2019}: morning, napper, afternoon, and evening. These clusters were identified in both studied contexts, despite the differences in how the courses were organized and, e.g., the amount of available daylight. It also seems that these distributions are not significantly influenced by flexible due dates or deadlines.

(RQ2) ``How do the typical working times of students relate to academic outcomes?'' We observed noticeable differences in the exam scores and project scores within the US context, where those active in the morning performed the best in the exam. No significant differences in exam scores or project scores were observed in the European context, although this could be partially influenced by a ceiling effect. In both contexts, the morning and napper chronotypes tended to start working on their projects earlier than the afternoon and evening chronotypes, and the napper chronotypes tended to type, in general, more than the other chronotypes.

(RQ3) ``How do contextual factors affect academic outcomes?'' We observed differences between the contexts in when the students started to work on their projects and noticed that, in general, students in the European context tended to start their work earlier. We hypothesized that this could be due to two factors; (1) using small exercises when starting with a new topic, and (2) having a strict no-late submissions policy. While we acknowledge that there are likely many other explanations for these observations, we posit that the way how courses are organized can lead to a situation where students are more likely to follow their diurnal preferences, i.e., work during times that are productive for them. 

In this article we have inferred behavior from observed keystrokes and while our conclusions are in line with prior theoretical and empirical research, we cannot for certain say whether our observations stem from students' circadian rhythms and diurnal preferences, or whether there are other factors at play. We do not know, for example, how many of the students have part-time jobs and how their working hours are distributed over the week. Regardless, we observe that the majority of the students tend to work during daylight hours, contrary to some of the stereotypes posited about CS students. And, more importantly, the chronotypes we discover and their correlations with academic outcomes align with prior studies, suggesting that we, as computing education researchers and practitioners, should take note and consider diurnal preference issues in course development and future research.

As a part of our future work, we are looking into the translation and use of a chronotype questionnaire (e.g. Caen Chronotype Questionnaire~\cite{dosseville2013validation}) and studying whether the chronotypes identified from keystroke data match those identified using a questionnaire. As previous research suggests that working on optimal hours of day contribute towards performance in complex tasks~\cite{Natale1997}, we are also looking into the interplay of task difficulty, chronotype, observed behavior, and performance, within the domain of learning programming.

\section{Data Availability}

Our human-centered study protocols prohibit us from making our keystroke data publicly available. Our code, however, is available at \url{http://doi.org/10.5281/zenodo.4498457}.

\balance

\bibliographystyle{IEEEtranN}

\begin{thebibliography}{45}
\providecommand{\natexlab}[1]{#1}
\providecommand{\url}[1]{#1}
\csname url@samestyle\endcsname
\providecommand{\newblock}{\relax}
\providecommand{\bibinfo}[2]{#2}
\providecommand{\BIBentrySTDinterwordspacing}{\spaceskip=0pt\relax}
\providecommand{\BIBentryALTinterwordstretchfactor}{4}
\providecommand{\BIBentryALTinterwordspacing}{\spaceskip=\fontdimen2\font plus
\BIBentryALTinterwordstretchfactor\fontdimen3\font minus
  \fontdimen4\font\relax}
\providecommand{\BIBforeignlanguage}[2]{{%
\expandafter\ifx\csname l@#1\endcsname\relax
\typeout{** WARNING: IEEEtranN.bst: No hyphenation pattern has been}%
\typeout{** loaded for the language `#1'. Using the pattern for}%
\typeout{** the default language instead.}%
\else
\language=\csname l@#1\endcsname
\fi
#2}}
\providecommand{\BIBdecl}{\relax}
\BIBdecl

\bibitem[Preckel et~al.(2011)Preckel, Lipnevich, Schneider, and
  Roberts]{preckel2011chronotype}
F.~Preckel, A.~A. Lipnevich, S.~Schneider, and R.~D. Roberts, ``Chronotype,
  cognitive abilities, and academic achievement: A meta-analytic
  investigation,'' \emph{Learning and Individual Differences}, vol.~21, no.~5,
  pp. 483--492, 2011.

\bibitem[Fucci et~al.(2020)Fucci, Scanniello, Romano, and Juristo]{Fucci2020}
\BIBentryALTinterwordspacing
D.~Fucci, G.~Scanniello, S.~Romano, and N.~Juristo, ``Need for sleep: The
  impact of a night of sleep deprivation on novice developers' performance,''
  \emph{{IEEE} Transactions on Software Engineering}, vol.~46, no.~1, pp.
  1--19, Jan. 2020. [Online]. Available:
  \url{https://doi.org/10.1109/tse.2018.2834900}
\BIBentrySTDinterwordspacing

\bibitem[Roenneberg et~al.(2003)Roenneberg, Wirz-Justice, and
  Merrow]{Roenneberg2003}
\BIBentryALTinterwordspacing
T.~Roenneberg, A.~Wirz-Justice, and M.~Merrow, ``Life between clocks: Daily
  temporal patterns of human chronotypes,'' \emph{Journal of Biological
  Rhythms}, vol.~18, no.~1, pp. 80--90, Feb. 2003. [Online]. Available:
  \url{https://doi.org/10.1177/0748730402239679}
\BIBentrySTDinterwordspacing

\bibitem[Leinonen et~al.(2016{\natexlab{a}})Leinonen, Longi, Klami, and
  Vihavainen]{leinonen2016automatic}
\BIBentryALTinterwordspacing
J.~Leinonen, K.~Longi, A.~Klami, and A.~Vihavainen, ``Automatic inference of
  programming performance and experience from typing patterns,'' in
  \emph{Proceedings of the 47th ACM Technical Symposium on Computing Science
  Education}, ser. SIGCSE '16.\hskip 1em plus 0.5em minus 0.4em\relax New York,
  NY, USA: ACM, 2016, pp. 132--137. [Online]. Available:
  \url{http://doi.acm.org/10.1145/2839509.2844612}
\BIBentrySTDinterwordspacing

\bibitem[Edwards et~al.(2020)Edwards, Leinonen, and Hellas]{edwards2020study}
J.~Edwards, J.~Leinonen, and A.~Hellas, ``A study of keystroke data in two
  contexts: Written language and programming language influence predictability
  of learning outcomes,'' in \emph{Proceedings of the 51st ACM Technical
  Symposium on Computer Science Education}, 2020, pp. 413--419.

\bibitem[Leinonen(2019)]{leinonen2019keystroke}
J.~Leinonen, ``Keystroke data in programming courses,'' Ph.D. dissertation,
  University of Helsinki, 2019.

\bibitem[Longi et~al.(2015)Longi, Leinonen, Nygren, Salmi, Klami, and
  Vihavainen]{longi2015identification}
K.~Longi, J.~Leinonen, H.~Nygren, J.~Salmi, A.~Klami, and A.~Vihavainen,
  ``Identification of programmers from typing patterns,'' in \emph{Proceedings
  of the 15th Koli Calling Conference on Computing Education Research}.\hskip
  1em plus 0.5em minus 0.4em\relax ACM, 2015, pp. 60--67.

\bibitem[Leinonen et~al.(2016{\natexlab{b}})Leinonen, Longi, Klami, Ahadi, and
  Vihavainen]{leinonen2016typing}
J.~Leinonen, K.~Longi, A.~Klami, A.~Ahadi, and A.~Vihavainen, ``Typing patterns
  and authentication in practical programming exams,'' in \emph{Proceedings of
  the 2016 ACM Conference on Innovation and Technology in Computer Science
  Education}.\hskip 1em plus 0.5em minus 0.4em\relax ACM, 2016, pp. 160--165.

\bibitem[Bergin and Reilly(2005)]{bergin2005influence}
S.~Bergin and R.~Reilly, ``The influence of motivation and comfort-level on
  learning to program,'' 2005.

\bibitem[Lepp{\"a}nen et~al.(2016)Lepp{\"a}nen, Leinonen, and
  Hellas]{leppanen2016pauses}
L.~Lepp{\"a}nen, J.~Leinonen, and A.~Hellas, ``Pauses and spacing in learning
  to program,'' in \emph{Proceedings of the 16th Koli Calling International
  Conference on Computing Education Research}, 2016, pp. 41--50.

\bibitem[Ilves et~al.(2018)Ilves, Leinonen, and Hellas]{ilves2018supporting}
K.~Ilves, J.~Leinonen, and A.~Hellas, ``Supporting self-regulated learning with
  visualizations in online learning environments,'' in \emph{Proceedings of the
  49th ACM Technical Symposium on Computer Science Education}, 2018, pp.
  257--262.

\bibitem[Edwards et~al.(2009)Edwards, Snyder, P{\'e}rez-Qui{\~n}ones, Allevato,
  Kim, and Tretola]{edwards2009comparing}
S.~H. Edwards, J.~Snyder, M.~A. P{\'e}rez-Qui{\~n}ones, A.~Allevato, D.~Kim,
  and B.~Tretola, ``Comparing effective and ineffective behaviors of student
  programmers,'' in \emph{Proceedings of the fifth international workshop on
  Computing education research workshop}, 2009, pp. 3--14.

\bibitem[Martin et~al.(2015)Martin, Edwards, and Shaffer]{martin2015effects}
J.~Martin, S.~H. Edwards, and C.~A. Shaffer, ``The effects of procrastination
  interventions on programming project success,'' in \emph{Proceedings of the
  eleventh annual International Conference on International Computing Education
  Research}, 2015, pp. 3--11.

\bibitem[Claes et~al.(2018)Claes, M\"{a}ntyl\"{a}, Kuutila, and
  Adams]{Claes2018}
\BIBentryALTinterwordspacing
M.~Claes, M.~V. M\"{a}ntyl\"{a}, M.~Kuutila, and B.~Adams, ``Do programmers
  work at night or during the weekend?'' in \emph{Proceedings of the 40th
  International Conference on Software Engineering}.\hskip 1em plus 0.5em minus
  0.4em\relax {ACM}, May 2018. [Online]. Available:
  \url{https://doi.org/10.1145/3180155.3180193}
\BIBentrySTDinterwordspacing

\bibitem[Reppert and Weaver(2002)]{Reppert2002}
\BIBentryALTinterwordspacing
S.~M. Reppert and D.~R. Weaver, ``Coordination of circadian timing in
  mammals,'' \emph{Nature}, vol. 418, no. 6901, pp. 935--941, Aug. 2002.
  [Online]. Available: \url{https://doi.org/10.1038/nature00965}
\BIBentrySTDinterwordspacing

\bibitem[Dibner et~al.(2010)Dibner, Schibler, and Albrecht]{Dibner2010}
\BIBentryALTinterwordspacing
C.~Dibner, U.~Schibler, and U.~Albrecht, ``The mammalian circadian timing
  system: Organization and coordination of central and peripheral clocks,''
  \emph{Annual Review of Physiology}, vol.~72, no.~1, pp. 517--549, Mar. 2010.
  [Online]. Available:
  \url{https://doi.org/10.1146/annurev-physiol-021909-135821}
\BIBentrySTDinterwordspacing

\bibitem[Bell-Pedersen et~al.(2005)Bell-Pedersen, Cassone, Earnest, Golden,
  Hardin, Thomas, and Zoran]{BellPedersen2005}
\BIBentryALTinterwordspacing
D.~Bell-Pedersen, V.~M. Cassone, D.~J. Earnest, S.~S. Golden, P.~E. Hardin,
  T.~L. Thomas, and M.~J. Zoran, ``Circadian rhythms from multiple oscillators:
  lessons from diverse organisms,'' \emph{Nature Reviews Genetics}, vol.~6,
  no.~7, pp. 544--556, Jul. 2005. [Online]. Available:
  \url{https://doi.org/10.1038/nrg1633}
\BIBentrySTDinterwordspacing

\bibitem[Wittmann et~al.(2006)Wittmann, Dinich, Merrow, and
  Roenneberg]{Wittmann2006}
\BIBentryALTinterwordspacing
M.~Wittmann, J.~Dinich, M.~Merrow, and T.~Roenneberg, ``Social jetlag:
  Misalignment of biological and social time,'' \emph{Chronobiology
  International}, vol.~23, no. 1-2, pp. 497--509, Jan. 2006. [Online].
  Available: \url{https://doi.org/10.1080/07420520500545979}
\BIBentrySTDinterwordspacing

\bibitem[Korczak et~al.(2008)Korczak, Martynhak, Pedrazzoli, Brito, and
  Louzada]{Korczak2008}
\BIBentryALTinterwordspacing
A.~Korczak, B.~Martynhak, M.~Pedrazzoli, A.~Brito, and F.~Louzada, ``Influence
  of chronotype and social zeitgebers on sleep/wake patterns,'' \emph{Brazilian
  Journal of Medical and Biological Research}, vol.~41, no.~10, pp. 914--919,
  Sep. 2008. [Online]. Available:
  \url{https://doi.org/10.1590/s0100-879x2008005000047}
\BIBentrySTDinterwordspacing

\bibitem[Vitale et~al.(2014)Vitale, Roveda, Montaruli, Galasso, Weydahl, Caumo,
  and Carandente]{Vitale2014}
\BIBentryALTinterwordspacing
J.~A. Vitale, E.~Roveda, A.~Montaruli, L.~Galasso, A.~Weydahl, A.~Caumo, and
  F.~Carandente, ``Chronotype influences activity circadian rhythm and sleep:
  Differences in sleep quality between weekdays and weekend,''
  \emph{Chronobiology International}, vol.~32, no.~3, pp. 405--415, Dec. 2014.
  [Online]. Available: \url{https://doi.org/10.3109/07420528.2014.986273}
\BIBentrySTDinterwordspacing

\bibitem[Valladares et~al.(2017)Valladares, Ram{\'{\i}}rez-Tagle, Mu{\~{n}}oz,
  and Obreg{\'{o}}n]{Valladares2017}
\BIBentryALTinterwordspacing
M.~Valladares, R.~Ram{\'{\i}}rez-Tagle, M.~A. Mu{\~{n}}oz, and A.~M.
  Obreg{\'{o}}n, ``Individual differences in chronotypes associated with
  academic performance among chilean university students,'' \emph{Chronobiology
  International}, vol.~35, no.~4, pp. 578--583, Dec. 2017. [Online]. Available:
  \url{https://doi.org/10.1080/07420528.2017.1413385}
\BIBentrySTDinterwordspacing

\bibitem[Porcheret et~al.(2018)Porcheret, Wald, Fritschi, Gerkema, Gordijn,
  Merrrow, Rajaratnam, Rock, Sletten, Warman, Wulff, Roenneberg, and
  Foster]{Porcheret2018}
\BIBentryALTinterwordspacing
K.~Porcheret, L.~Wald, L.~Fritschi, M.~Gerkema, M.~Gordijn, M.~Merrrow, S.~M.
  Rajaratnam, D.~Rock, T.~L. Sletten, G.~Warman, K.~Wulff, T.~Roenneberg, and
  R.~G. Foster, ``Chronotype and environmental light exposure in a student
  population,'' \emph{Chronobiology International}, vol.~35, no.~10, pp.
  1365--1374, Jun. 2018. [Online]. Available:
  \url{https://doi.org/10.1080/07420528.2018.1482556}
\BIBentrySTDinterwordspacing

\bibitem[Putilov et~al.(2019)Putilov, Marcoen, Neu, Pattyn, and
  Mairesse]{Putilov2019}
\BIBentryALTinterwordspacing
A.~A. Putilov, N.~Marcoen, D.~Neu, N.~Pattyn, and O.~Mairesse, ``There is more
  to chronotypes than evening and morning types: Results of a large-scale
  community survey provide evidence for high prevalence of two further types,''
  \emph{Personality and Individual Differences}, vol. 148, pp. 77--84, Oct.
  2019. [Online]. Available: \url{https://doi.org/10.1016/j.paid.2019.05.017}
\BIBentrySTDinterwordspacing

\bibitem[Kolomeichuk et~al.(2016)Kolomeichuk, Randler, Shabalina, Fradkova, and
  Borisenkov]{Kolomeichuk2016}
\BIBentryALTinterwordspacing
S.~N. Kolomeichuk, C.~Randler, I.~Shabalina, L.~Fradkova, and M.~Borisenkov,
  ``The influence of chronotype on the academic achievement of children and
  adolescents {\textendash} evidence from russian karelia,'' \emph{Biological
  Rhythm Research}, vol.~47, no.~6, pp. 873--883, Jul. 2016. [Online].
  Available: \url{https://doi.org/10.1080/09291016.2016.1207352}
\BIBentrySTDinterwordspacing

\bibitem[Montaruli et~al.(2019)Montaruli, Castelli, Galasso, Mul{\`{e}}, Bruno,
  Esposito, Caumo, and Roveda]{Montaruli2019}
\BIBentryALTinterwordspacing
A.~Montaruli, L.~Castelli, L.~Galasso, A.~Mul{\`{e}}, E.~Bruno, F.~Esposito,
  A.~Caumo, and E.~Roveda, ``Effect of chronotype on academic achievement in a
  sample of italian university students,'' \emph{Chronobiology International},
  vol.~36, no.~11, pp. 1482--1495, Aug. 2019. [Online]. Available:
  \url{https://doi.org/10.1080/07420528.2019.1652831}
\BIBentrySTDinterwordspacing

\bibitem[Dunn(1987)]{dunn1987research}
R.~Dunn, ``Research on instructional environments: Implications for student
  achievement and attitudes.'' \emph{Professional School Psychology}, vol.~2,
  no.~1, p.~43, 1987.

\bibitem[Dunn and Griggs(1988)]{dunn1988learning}
R.~Dunn and S.~A. Griggs, \emph{Learning styles: Quiet revolution in American
  secondary schools.}\hskip 1em plus 0.5em minus 0.4em\relax ERIC, 1988.

\bibitem[Nowack and Van Der~Meer(2018)]{nowack2018synchrony}
K.~Nowack and E.~Van Der~Meer, ``The synchrony effect revisited: chronotype,
  time of day and cognitive performance in a semantic analogy task,''
  \emph{Chronobiology international}, vol.~35, no.~12, pp. 1647--1662, 2018.

\bibitem[Cajochen et~al.(1999)Cajochen, Khalsa, Wyatt, Czeisler, and
  Dijk]{Cajochen1999}
\BIBentryALTinterwordspacing
C.~Cajochen, S.~B.~S. Khalsa, J.~K. Wyatt, C.~A. Czeisler, and D.-J. Dijk,
  ``{EEG} and ocular correlates of circadian melatonin phase and human
  performance decrements during sleep loss,'' \emph{American Journal of
  Physiology-Regulatory, Integrative and Comparative Physiology}, vol. 277,
  no.~3, pp. R640--R649, Sep. 1999. [Online]. Available:
  \url{https://doi.org/10.1152/ajpregu.1999.277.3.r640}
\BIBentrySTDinterwordspacing

\bibitem[Schmidt et~al.(2007)Schmidt, Collette, Cajochen, and
  Peigneux]{Schmidt2007}
\BIBentryALTinterwordspacing
C.~Schmidt, F.~Collette, C.~Cajochen, and P.~Peigneux, ``A time to think:
  Circadian rhythms in human cognition,'' \emph{Cognitive Neuropsychology},
  vol.~24, no.~7, pp. 755--789, Oct. 2007. [Online]. Available:
  \url{https://doi.org/10.1080/02643290701754158}
\BIBentrySTDinterwordspacing

\bibitem[Valdez et~al.(2008)Valdez, Reilly, and Waterhouse]{Valdez2008}
\BIBentryALTinterwordspacing
P.~Valdez, T.~Reilly, and J.~Waterhouse, ``Rhythms of mental performance,''
  \emph{Mind, Brain, and Education}, vol.~2, no.~1, pp. 7--16, Mar. 2008.
  [Online]. Available: \url{https://doi.org/10.1111/j.1751-228x.2008.00023.x}
\BIBentrySTDinterwordspacing

\bibitem[Hasher et~al.(2008)Hasher, Lustig, and Zacks]{Hasher2008}
\BIBentryALTinterwordspacing
L.~Hasher, C.~Lustig, and R.~Zacks, ``Inhibitory mechanisms and the control of
  attention,'' in \emph{Variation in Working Memory}.\hskip 1em plus 0.5em
  minus 0.4em\relax Oxford University Press, Mar. 2008, pp. 227--249. [Online].
  Available: \url{https://doi.org/10.1093/acprof:oso/9780195168648.003.0009}
\BIBentrySTDinterwordspacing

\bibitem[Natale and Lorenzetti(1997)]{Natale1997}
\BIBentryALTinterwordspacing
V.~Natale and R.~Lorenzetti, ``Influences of morningness-eveningness and time
  of day on narrative comprehension,'' \emph{Personality and Individual
  Differences}, vol.~23, no.~4, pp. 685--690, Oct. 1997. [Online]. Available:
  \url{https://doi.org/10.1016/s0191-8869(97)00059-7}
\BIBentrySTDinterwordspacing

\bibitem[Natale et~al.(2003)Natale, Alzani, and Cicogna]{Natale2003}
\BIBentryALTinterwordspacing
V.~Natale, A.~Alzani, and P.~Cicogna, ``Cognitive efficiency and circadian
  typologies: a diurnal study,'' \emph{Personality and Individual Differences},
  vol.~35, no.~5, pp. 1089--1105, Oct. 2003. [Online]. Available:
  \url{https://doi.org/10.1016/s0191-8869(02)00320-3}
\BIBentrySTDinterwordspacing

\bibitem[Leinonen et~al.(2017)Leinonen, Lepp{\"a}nen, Ihantola, and
  Hellas]{leinonen2017comparison}
J.~Leinonen, L.~Lepp{\"a}nen, P.~Ihantola, and A.~Hellas, ``Comparison of time
  metrics in programming,'' in \emph{Proceedings of the 2017 acm conference on
  international computing education research}, 2017, pp. 200--208.

\bibitem[Pintrich et~al.(1991)]{pintrich1991manual}
P.~R. Pintrich \emph{et~al.}, ``A manual for the use of the motivated
  strategies for learning questionnaire ({MSLQ}).'' 1991.

\bibitem[Spacco et~al.(2015)Spacco, Denny, Richards, Babcock, Hovemeyer,
  Moscola, and Duvall]{spacco2015analyzing}
J.~Spacco, P.~Denny, B.~Richards, D.~Babcock, D.~Hovemeyer, J.~Moscola, and
  R.~Duvall, ``Analyzing student work patterns using programming exercise
  data,'' in \emph{Proceedings of the 46th ACM Technical Symposium on Computer
  Science Education}, 2015, pp. 18--23.

\bibitem[Auvinen et~al.(2015)Auvinen, Hakulinen, and
  Malmi]{auvinen2015increasing}
T.~Auvinen, L.~Hakulinen, and L.~Malmi, ``Increasing students’ awareness of
  their behavior in online learning environments with visualizations and
  achievement badges,'' \emph{IEEE Transactions on Learning Technologies},
  vol.~8, no.~3, pp. 261--273, 2015.

\bibitem[Denny et~al.(2018)Denny, Luxton-Reilly, Craig, and
  Petersen]{denny2018improving}
P.~Denny, A.~Luxton-Reilly, M.~Craig, and A.~Petersen, ``Improving complex task
  performance using a sequence of simple practice tasks,'' in \emph{Proceedings
  of the 23rd Annual ACM Conference on Innovation and Technology in Computer
  Science Education}, 2018, pp. 4--9.

\bibitem[Prather et~al.(2020)Prather, Becker, Craig, Denny, Loksa, and
  Margulieux]{prather2020we}
J.~Prather, B.~A. Becker, M.~Craig, P.~Denny, D.~Loksa, and L.~Margulieux,
  ``What do we think we think we are doing? metacognition and self-regulation
  in programming,'' in \emph{Proceedings of the 2020 ACM Conference on
  International Computing Education Research}, 2020, pp. 2--13.

\bibitem[Margulieux et~al.(2012)Margulieux, Guzdial, and
  Catrambone]{margulieux2012subgoal}
\BIBentryALTinterwordspacing
L.~E. Margulieux, M.~Guzdial, and R.~Catrambone, ``Subgoal-labeled
  instructional material improves performance and transfer in learning to
  develop mobile applications,'' in \emph{Proceedings of the Ninth Annual
  International Conference on International Computing Education Research}, ser.
  ICER '12.\hskip 1em plus 0.5em minus 0.4em\relax New York, NY, USA:
  Association for Computing Machinery, 2012, p. 71–78. [Online]. Available:
  \url{https://doi.org/10.1145/2361276.2361291}
\BIBentrySTDinterwordspacing

\bibitem[Hellas et~al.(2017)Hellas, Leinonen, and
  Ihantola]{hellas2017plagiarism}
A.~Hellas, J.~Leinonen, and P.~Ihantola, ``Plagiarism in take-home exams:
  help-seeking, collaboration, and systematic cheating,'' in \emph{Proceedings
  of the 2017 ACM conference on innovation and technology in computer science
  education}, 2017, pp. 238--243.

\bibitem[Wasserstein and Lazar(2016)]{wasserstein2016asa}
R.~L. Wasserstein and N.~A. Lazar, ``The asa statement on p-values: context,
  process, and purpose,'' 2016.

\bibitem[Carver(1993)]{carver1993case}
R.~P. Carver, ``The case against statistical significance testing, revisited,''
  \emph{The Journal of Experimental Education}, vol.~61, no.~4, pp. 287--292,
  1993.

\bibitem[Dosseville et~al.(2013)Dosseville, Laborde, and
  Lericollais]{dosseville2013validation}
F.~Dosseville, S.~Laborde, and R.~Lericollais, ``Validation of a chronotype
  questionnaire including an amplitude dimension,'' \emph{Chronobiology
  international}, vol.~30, no.~5, pp. 639--648, 2013.

\end{thebibliography}

\end{document}